\documentclass{aa} 
\usepackage{tikz}
\usepackage{amsmath}
\usepackage{graphicx}
\usepackage{mwe}
\usepackage{graphicx}

\usepackage{txfonts}

\usepackage{xcolor}
\usepackage{comment}
\usepackage{url}

\usepackage{caption}
\usepackage{subcaption}
\newcommand{\IH}[1]{{\color{blue}[Ian: #1]}}

\usepackage{hyperref} 
\begin{document}

   \title{Generative Models of 21cm EoR Lightcones with 3D Scattering Transforms}
\titlerunning{3D Scattering Transforms}

   \author{Ian Hothi 
     \inst{1,}\inst{2}\fnmsep\thanks{Ian.Hothi@obspm.fr}
     \and
     Erwan Allys\inst{1}
     \and
     Benoit Semelin\inst{2}
     \and
     Romain Meriot\inst{3}
     }

  \institute{
   Laboratoire de Physique de l’ENS, ENS, Universit\'{e} PSL, CNRS, Sorbonne Universit\'{e}, Universit\'{e}e Paris Cit\'{e}, 75005 Paris, France
     \and
      LERMA, Observatoire de Paris, PSL Research University, CNRS, Sorbonne Universit\'{e}, F-75014 Paris, France
            \and
      Department of Physics, Blackett Laboratory, Imperial College London, SW7 2AZ, U.K
       }

  \date{Received ; accepted }

% \abstract{}{}{}{}{} 
% 5 {} token are mandatory

  \abstract
  {The 21cm signal from the Epoch of Reionization (EoR) is observed as a three-dimensional data set known as a lightcone, consisting of a redshift (frequency) axis and two spatial sky plane axes. When observed by radio interferometers, this EoR signal is strongly obscured by foregrounds that are several orders of magnitude stronger. Due to its inherently non-Gaussian nature, the EoR signal requires robust statistical tools to accurately separate it from these foreground contaminants, but current foreground separation techniques focus primarily on recovering the EoR power spectrum, often neglecting valuable non-Gaussian information. Recent developments in astrophysics, particularly in the context of the Galactic interstellar medium, have demonstrated the efficacy of scattering transforms - novel summary statistics for highly non-Gaussian processes - for component separation tasks. Motivated by these advances, we extend the scattering transform formalism from two-dimensional data sets to three-dimensional EoR lightcones. To this end, we introduce a 3D wavelet set from the tensor product of 2D isotropic wavelets in the sky plane domain and 1D wavelets in the redshift domain. As generative models form the basis of component separation, our focus here is on building and validating generative models that can be used for component separation in future projects. To achieve this, we construct maximum entropy generative models to synthesise EoR lightcones, and statistically validate the generative model by quantitatively comparing the synthesised EoR lightcones with the single target lightcone used to construct them, using independent statistics such as the power spectrum and Minkowski Functionals. The synthesised lightcones agree well with the target lightcone both statistically and visually, opening up the possibility of developing for component separation methods using 3D scattering transforms.}
  \keywords{Cosmology:dark ages,reionization,first stars – Early Universe – Methods:statistical}

   \maketitle
%
%-------------------------------------------------------------------

\section{Introduction}
The Epoch of Reionization (EoR) is a critical transition in the history of the Universe, marking the emergence of the first stars and the eventual ionization of the once-neutral Intergalactic Medium (IGM). Indirect observations of the EoR from the Cosmic Microwave Background (CMB) \citep{PlanckEoR16, 2018:GorceDouspisAghanim} and Gunn-Peterson trough in Quasar spectra \citep{GPTroughFan,2015:BeckerBoltonMadau, 2018:BosmanFanJiang, 2021:QinMesingerBosman}, provide constraints on the timings of the EoR to be between redshifts $ 14 \gtrsim  z \gtrsim 5.3$. While these offer loose timings of the EoR, these indirect observations do not constrain the sources responsible for the EoR nor the state of the IGM, for which a direct probe is needed. The redshifted 21cm signal, resulting from the hyperfine transition in neutral hydrogen during the EoR, presents a promising direct observational probe (observed by radio interferometers between $\sim$ 100MHz to 200 MHz) for a comprehensive study of this epoch. 

Observations of the 21cm signal (EoR signal, henceforth) by radio interferometers must contend with foregrounds such as synchrotron emission from diffuse clouds in the Galaxy, noise arising from the antennas, and systematics such as mode-mixing arising from the chromatic nature of radio interferometers. All of these contaminants are orders of magnitude stronger in total variance than the signal itself \cite{2008:JelicZaroubiLabropoulos,2021:HothiChapmanPritchard}. 
\begin{figure}
\centering
 \includegraphics[width=.75\linewidth]{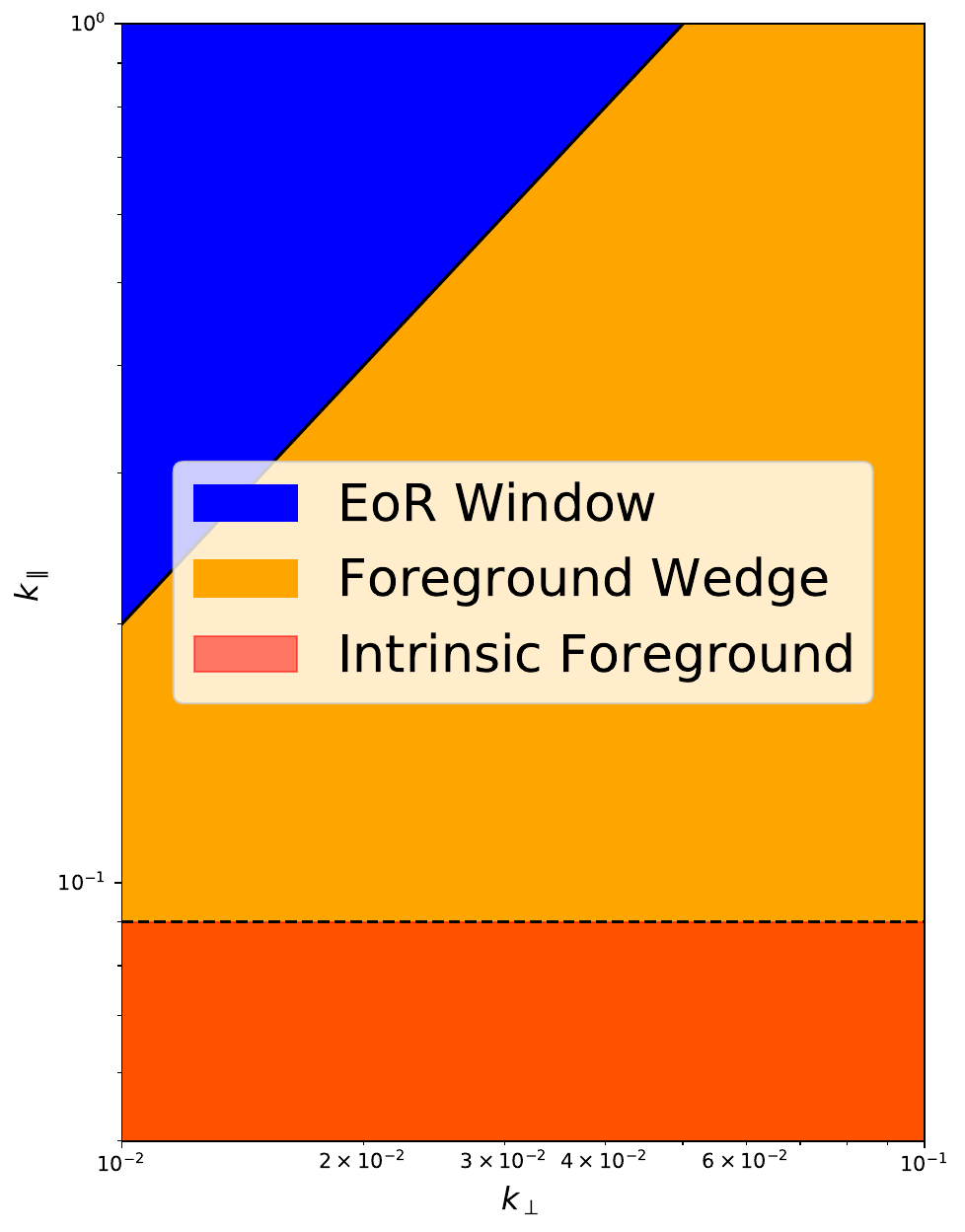}
\caption{Schematic of the cylindrically-averaged power spectrum of an observed EoR Lightcone, with the three distinct regions highlighted.}
\label{fig:EoR_Window}
\end{figure}
The cylindrically-averaged power spectrum of a radio interferometer observation, Fig. \ref{fig:EoR_Window}, has three distinct regions in the $k_\parallel \; vs. \; k_\perp$ plane\footnote{ The $z$-domain, or line-of-sight, will be denoted by $\parallel$, and the $xy$-plane, the perpendicular sky domain, denoted by $\perp$.}.
The first region, located at low $k_{\parallel}$, is where the astrophysical foregrounds dominate. These astrophysical foregrounds are made up of the Galactic emission, arising from free-free bremsstrahlung emission and diffuse synchrotron emission, as well as extragalactic synchrotron emission arising from sources such as radio halos and clusters \citep{2008:JelicZaroubiLabropoulos,2021:HothiChapmanPritchard}. 
The second region is the foreground wedge, which arises due to the chromatic nature of the instrument, the fact that the interferometer's point spread function (PSF) is frequency dependent, leading to mode mixing between $k_{\perp}$ and $k_{\parallel}$, and causing the foreground emission to leak to higher $k_{\parallel}$\citep{2014:LiuParsonsTrott}. Bright sources in the sky that come into one of the side lobes of the interferometers PSF also contribute to the emission within the wedge - all of which are bound by the horizon\footnote{However, foregrounds can leak across the wedge line \citep{2018:MurrayTrott}.}.
The third region is the EoR window, where the foregrounds and instrumental effects are subdominant (there is little to no contribution on these scales) to the noise and EoR signal.

There are two schools of mitigation of these foregrounds: Avoidance and Removal. Avoidance consists in focusing on the EoR window, where the foregrounds and systematics are subdominant to the EoR signal and noise~\citep{2014:LiuParsonsTrott}. In contrast, removal attempts to model the foregrounds and systematics and subtract them from the observed signal, leaving a residual signal free of the contaminants. Since contaminants models are usually defined over the entire Fourier space, removal approaches can extract information from Fourier modes outside the EoR window, which is not possible with avoidance. In particular, early removal techniques, such as polynomial fitting~\citep{2008:JelicZaroubiLabropoulos}, took advantage of the smooth nature of radio foregrounds in the frequency domain. However, instrumental effects can affect this smoothness, which led to the development of non-parametric methods such as Blind Source Separation (BSS), which do not assume a predefined parametrisation of the foregrounds \citep{FASTICA12a,GMCA13}. Gaussian Process Regression (GPR) has also emerged as an optimised fitting tool, using priors on the frequency covariance of various components of the observed data, including foregrounds and the EoR signal \cite{2024:AcharyaMertensCiardi}.

These current Removal techniques have been optimised to recover the power spectrum of the EoR signal. However, the EoR signal is expected to be highly non-Gaussian, and this non-Gaussian information is important to constrain not only the astrophysics of this epoch, but also the progression of reionization itself \cite{2017:ShimabukuroYoshiuraTakahashi,2022:WatkinsonGreigMesinger,2022:TiwariShawMajumdar}. To develop a removal/component separation algorithm that is efficient at statistically recovering the EoR signal beyond the power spectrum, we propose in this paper to rely on beyond power spectrum statistics, such as Scattering Transform statistics, to perform the separation itself.

Scattering transforms (ST) have emerged as a novel set of summary statistics, which are constructed by successive wavelet convolutions and the application of non-linear operators such as modulus \citep{2011:Mallat,2012:BrunaMallat}. They have arisen as powerful low-variance statistics to characterise non-Gaussian processes, and have already been utilised with success in a range of topics, including the study of the Interstellar medium \citep{2019:AllysLevrierZhang,regaldo2020statistical,lei2023probing}; weak lensing~\citep{2020:ChengTingMenard}; Large-scale structure of the Universe for 2D~\citep{2020:AllysMarchandCardoso} and 3D data~\citep{2022:EickenbergAllysMoradinezhadDizgah,valogiannis2022towards,2024:Regaldo-SaintBlancardHahnHo}; and more recently to the EoR \citep{2022:GreigTingKaurov,2024:ZhaoMaoZuo,2023:HothiAllysSemelin,2024:PrelogovicMesinger,2025:ShimabukuroXuShao}. 

An advantage of ST statistics is that they can be used to build generative models of physical processes, in the framework of maximum entropy models. Such models, parametrised by the ST statistics themselves, have been shown to be very efficient in approximating complex non-Gaussian physical fields, even when constructed from a single sample~\cite{2020:AllysMarchandCardoso,2023:ChengMorelAllys,2024:MoussetAllysPrice}. ST have also been used in the context of EoR to build diffusion models, using U-Nets, and have been shown to outperform Generative adversarial networks when applied to 2D EoR fields \citep{2023:ZhaoTingDiao}.

The ability to generate realistic realisations of various processes under ST constraints led to the development of new component separation algorithms. These ST-based component separations have already been successfully applied directly on observational data, relying on the non-Gaussian properties of the components within the data, without physically driven priors on the component of interest \citep{2021:Regaldo-SaintBlancardAllysBoulanger,2022:DelouisAllysGauvrit,2024:AuclairAllysBoulanger}. However, these previous algorithms have only been applied to 2D data. For EoR studies, these component separations must be extended to 3D EoR lightcones. This in turn requires reliance on ST statistics that can adequately describe such data, the development of which is the purpose of this paper. To this end, this work extends ST generative models to 3-dimensional EoR data and quantitatively validates the quality of such models.

To construct such 3D ST statistics, a first step is to introduce a relevant set of wavelets that characterise the different regions of the $k_\parallel \; vs. \; k_\perp$ plane, see~Fig.~\ref{fig:EoR_Window}. In fact, the wavelet sets used in previous 3D ST applications usually probe extended regions of the $k_\parallel \; vs. \; k_\perp$ plane, thus mixing regions where the foregrounds are dominant and sub-dominant~\citep[see, for example,][]{2022:EickenbergAllysMoradinezhadDizgah}. This is problematic because mixing regions where foregrounds are dominant with regions where they are subdominant is detrimental to the efficient extraction of information from the EoR signal. To avoid this drawback, we construct in this paper a new set of 3D wavelets from the tensor product between a set of 1D wavelets along the $k_\parallel$ axis and a set of 2D isotropic wavelets along the sky axes ($k_x,k_y$). This new set of wavelets is then used to construct 3D ST statistics, which in turn are used to construct a generative model of EoR lightcones, the quality of which is then assessed quantitatively. To do this, such an ST model is built from a single simulated EoR lightcone, whose statistical properties are compared with those of the simulation using several independent and complementary statistics. 

The paper is structured as follows: The simulations used in this work are outlined in Section \ref{simulation}. Section \ref{Development} presents the 3D scattering transforms. The generative model built from scattering transforms are then described in Section \ref{Method}. Finally, the comparison of the synthesised lightcones to the target EoR lightcones is presented in Section \ref{Results}.

%%%%%%%%%%%%%%%%%%%%%%%%%%%%%%%%%%%%%%%%%%%%%%%%%%%%%%%%%%%%%%%%%%%%%%%%%%%%%%
%%%%%%%%%%%%%%%%%%%%%%%%%%%%%%%%%%%%%%%%%%%%%%%%%%%%%%%%%%%%%%%%%%%%%%%%%%%%%%
\section{Simulation: \textit{LoReLi II}}
\label{simulation}

The simulation used in this work is from the \textit{LoReLi II} (LOw REsolution LIcorice) database \citep{2023:MeriotSemelin,2024:MeriotSemelin1}. This database is a collection of Licorice simulations designed to study the EoR signal~\citep{2007:SemelinCombesBaek,2017:SemelinEamesBolgar}

The \textit{LoReLi II} database comprises 9828 simulations of the EoR signal's brightness temperature, and explores a five-parameter space, which are the gas-to-star conversion timescale, minimum halo mass for star formation, escape fraction of UV radiation, and X-ray production efficiency, while keeping cosmological parameters constant at $\{$ $H_0$: 67.8 km/s/Mpc, $\Omega_0$: 0.308, $\Omega_b$: 0.0484, $\Omega_{\Lambda}$: 0.692, $\sigma_8$:0.815 $\}$. This setup allows for a detailed exploration of the astrophysical processes during reionization without the confounding effects of varying cosmology.

In this paper, the ST generative model is built from a single \textit{LoReLi} simulation of a lightcone. The astrophysical parameters of the lightcone, which we refer to as the target lightcone, are:
\begin{itemize}
    \item log10 of minimum mass of halos, $M_{min}$: 8.0
    \item X ray luminosity fraction, $f_x$: 0.1
    \item Hard X ray fraction, $r_{\frac{H}{S}}$: 0.2
    \item Gas conversion time scale for star formation in 10 Myr, $\tau$:~8191.9
    \item Ionizing escape fraction, $f_{esc,post}$: 0.275.
\end{itemize}

This lightcone has a redshift range between $23 \leq z \leq 5$. For the $xy$-plane, the 2D sky, there are $256^2$ pixels and for the $z$-domain there are 3072 redshift channels between $z = 23$ and $z = 5$. To reduce the computational cost and to limit the $z$ evolution over the lightcone, we downgrade it from $256^2$ pixels to $128^2$ pixels in the $xy$ plane, and we limit the $z$ values between $9.5 \leq z \leq 8.5$ (frequency bandwidth of $\delta \nu \sim 14 MHz$), corresponding to 170 pixels in the $z$ domain, for a final data cube of (128,128,170) pixels. Indeed, if the ST statistics were estimated on the initial lightcone, the statistical properties would evolve with $z$, whereas we approximate that there is only a limited $z$ evolution with the final constructed lightcone. We note, however, that a lightcone with a larger redshift range could be built by stacking models constructed with a more limited redshift range.

%%%%%%%%%%%%%%%%%%%%%%%%%%%%%%%%%%%%%%%%%%%%%%%%%%%%%%%%%%%%%%%%%%%%%%%%%%%%%%
%%%%%%%%%%%%%%%%%%%%%%%%%%%%%%%%%%%%%%%%%%%%%%%%%%%%%%%%%%%%%%%%%%%%%%%%%%%%%%
\section{Scattering Transforms for 3D Lightcones}
\label{Development}
%%%%%%%%%%%%%%%%%%%%%%%%%%%%%%%%%%%%%%%%%%%%%%%%%%%%%%%%%%%%%%%%%%%%%%%%%%%%%%
\subsection{Wavelets for 3D Lightcones}
\label{PartWavelets}

\begin{figure}
\vspace{-0.6cm}
\centering
{%
\begin{tikzpicture}
    \node (image) at (0,0) {%
        \includegraphics[width=.9\linewidth]{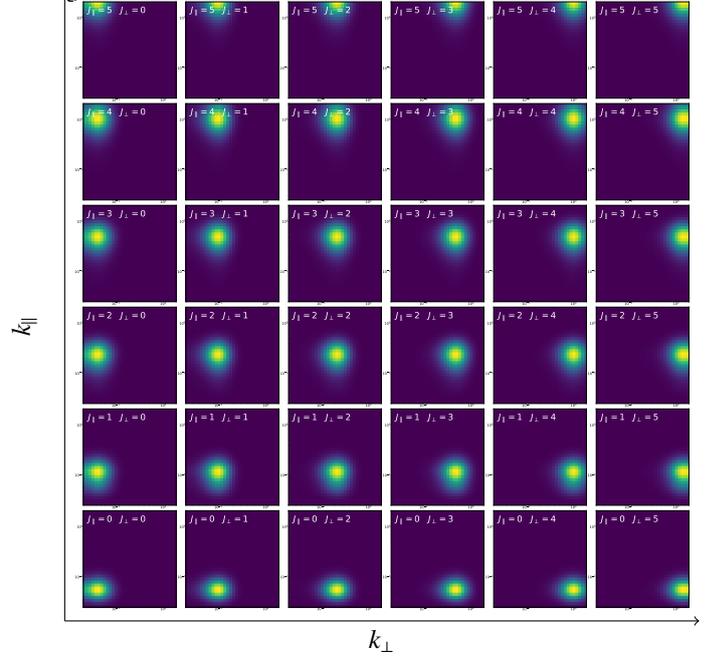}%
    };

    % X-axis arrow and label (bottom edge)
    \draw [->] (image.south west) -- (image.south east)
        node[midway,below] {$k_{\perp}$};

    % Y-axis arrow and label (left edge); rotate label 90 deg if you prefer
    \draw [->] (image.south west) -- (image.north west)
        node[midway,left=15pt,rotate=90] {$k_{\parallel}$};

\end{tikzpicture}} 
\caption{The different regions in the $k_\parallel$ vs. $k_\perp$ plane that the wavelets in the 3D wavelet set characterise.}
\label{fig:Fourier_Wavelet_Space}
\end{figure}

In this paper, we introduce a wavelet set that characterises the $k_\parallel \; vs. \; k_\perp$ plane by performing a tensor product between wavelet sets defined on different axes. The first is a 1D wavelet set defined in $k_z$, and denoted by $\parallel$, and the second is a 2D isotropic wavelet set defined in the $k_x-k_y$ plane, and denoted by $\perp$. To simplify notations, we also define $k_\parallel = k_z$ and $k_\perp = \sqrt{k^2_x+k^2_y}$. 

The mother wavelet of the 1D $\parallel$ wavelet set, which is $L_2$-normalised, is defined as:
\begin{equation}
\label{EqMotherParallel}
    \psi^{\parallel}_0({k_z}) = \psi^{\parallel}_0({k_{\parallel}}) = \frac{1}{\sqrt{2\pi\sigma_0^{\parallel\;2}}}\exp{-\frac{\left(|k_\parallel|-\mu^{\parallel}_0\right)^2}{2\sigma_0^{\parallel\;2}}},
\end{equation}
where $\mu^{\parallel}_0$ is the central wavenumber of the mother wavelet in Fourier space, and $\sigma^{\parallel}_0$ its width. The mother wavelet of the 2D $\perp$ isotropic wavelet set, which is $L_2$-normalised, is defined as: 
\begin{equation}
\label{EqMotherPerp}
    \psi^{\perp}_0(k_{x},k_{y}) = \psi^{\perp}_0({k_\perp}) = \frac{1}{{2\pi\sigma_0^{\perp\;2}}}\exp{-\frac{\left(|k_\perp|-\mu^{\perp}_0\right)^2}{2\sigma_0^{\perp\;2}}},
\end{equation}
where $\mu^{\perp}_0$ is again the central wavenumber of the mother wavelet in Fourier space, and $\sigma^{\perp}_0$ its width.

The two wavelet sets are then constructing by dilating the respective mother wavelets 
\begin{align}
        \psi^{\parallel}_{j_{\parallel}}(k_\parallel) &= 2 ^{-\frac{1}{2}\frac{j_{\parallel}}{Q_{\parallel}}}\psi^{\parallel}_0 \left(2^{-\frac{j_{\parallel}}{Qj_{\parallel}}}k_\parallel\right),
        \\
        \psi^{\perp}_{j_{\perp}}(k_\perp) &= 2 ^{-\frac{1}{2}\frac{2j_{\perp}}{Qj_{\perp}}}\psi^{\perp}_0 \left(2^{-\frac{j_{\perp}}{Q_{\perp}}}k_\perp\right),
\end{align}
where, $j_\parallel$ and $j_\perp$ are the integer scales of the dilation, and $Q_\parallel$ and $Q_\perp$ are the global quality factors. From these equations, one can check that the $\psi^{\parallel}_{j_{\parallel}}(k_z)$ and $\psi^{\perp}_{j_{\perp}}(k_\perp)$ can be written in the same form as the mother wavelets defined Eqs.~\eqref{EqMotherParallel} and~\eqref{EqMotherPerp}, by making the following replacements:
\begin{equation}
\begin{array}{l}
     \sigma^{\alpha}_0 \rightarrow \sigma^\alpha_{j_\alpha} = 2^{\frac{j_\alpha}{Q_\alpha}}\sigma^\alpha_0,  \\
     \mu^{\alpha}_0 \rightarrow \mu^\alpha_{j_\alpha} = 2^{\frac{j_\alpha}{Q_\alpha}}\mu^\alpha_0,
\end{array}
\end{equation}
for $\alpha \in \{\parallel,\perp\}$. The $j_\alpha$ integer scales of the $\psi_{j_\alpha}$ wavelets range between 0 and $J_\alpha - 1$, where $J_\alpha$ is the number of scales. Once $J_\alpha$ is chosen, the value of $Q_\alpha$ is fixed so that the wavelet set samples the full extent of the Fourier space. Specifically, when the mother wavelet, centred at $\mu^\alpha_{0}$, is dilated up to the largest scale, the resulting $\psi_{J_\alpha-1}$ wavelet is centred at the maximum Fourier scale $k_{\alpha,max}$, \textit{i.e.}, the Nyquist scale, yielding:
\begin{equation}
\label{get_Q}
    \log{\frac{k_{\alpha,max}}{\mu^\alpha_{0}}} = \frac{J_\alpha-1}{Q_\alpha} \log{2}.
\end{equation}
To construct the 3D wavelet, we consider the tensor product between the two wavelet sets we have previously defined: 
\begin{equation}
\Psi_{j_\parallel,j_\perp}(k_\parallel,k_\perp) = \psi^{\parallel}_{j_{\parallel}}(k_\parallel)\otimes \psi^{\perp}_{j_{\perp}}(k_\perp) = \psi^{\parallel}_{j_{\parallel}}(k_\parallel) \cdot \psi^{\perp}_{j_{\perp}}(k_\perp),
\end{equation}
where $\otimes$ is the tensor product. Also, in the following, we use the shorthand notation $\lambda = \{j_\parallel,j_\perp\}$, yielding
\begin{equation}
\Psi_{\lambda}(k_\parallel,k_\perp) = \Psi_{j_\parallel,j_\perp}(k_\parallel,k_\perp).
\end{equation}
These wavelets are real in Fourier space, and complex in real space.

In this paper, the central frequencies $\{\mu^\parallel_0,\mu^\perp_0\}$ of the mother wavelets were chosen to be two times the smallest Fourier scale available, giving $\{\mu^\parallel_0,\mu^\perp_0\} = \{0.023 \text{ hMpc}^{-1}, 0.028 \text{ hMpc}^{-1}\}$. This allows sampling on scales as large as possible, while still being able to estimate statistics on non-periodic fields (see Sec.~\ref{Part_NonPBC} below). We then chose $(J_\parallel,J_\perp) = (6,6)$, corresponding to $(Q_\parallel,Q_\perp) = (0.90,0.93)$, which is close to the commonly used dyadic scaling, resulting in a total of 36 wavelets. The values $\{\sigma^\parallel_0,\sigma^\perp_0\} = \{0.011 \text{ hMpc}^{-1}, 0.015 \text{ hMpc}^{-1}\}$ were finally chosen in order to cover the entire $k_\parallel \; vs. \; k_\perp$ Fourier plane as uniformly as possible. 
A plot of the resulting set of wavelets in this cylindrical Fourier space is shown in Fig.~\ref{fig:Fourier_Wavelet_Space}.

%%%%%%%%%%%%%%%%%%%%%%%%%%%%%%%%%%%%%%%%%%%%%%%%%%%%%%%%%%%%%%%%%%%%%%%%%%%%%%
\subsection{Scattering Transform Statistics}
\label{WSS}

\begin{figure}
\centering
\includegraphics[width=.75\linewidth]{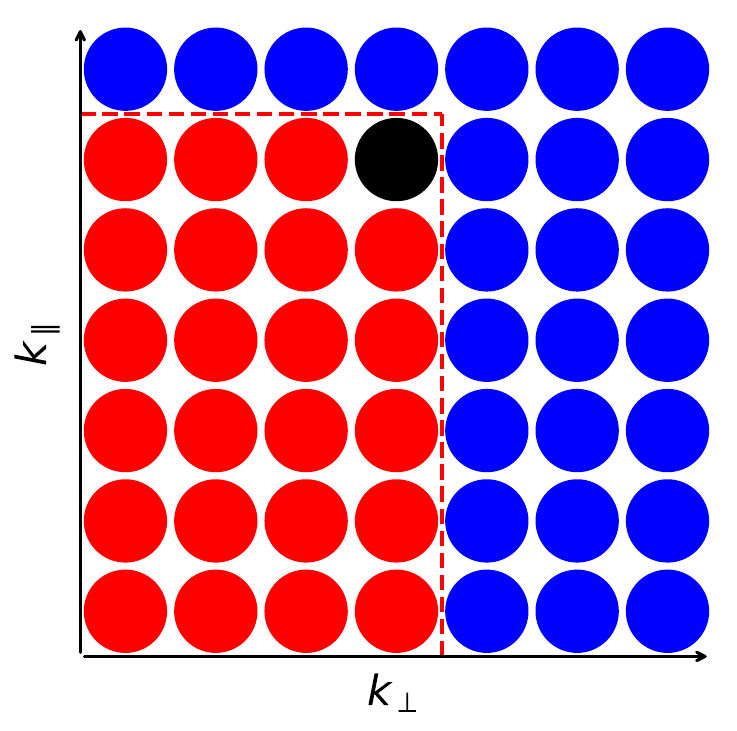}
\caption{Schematic of the secondary convolution criterion $\lambda_1 \leq \lambda_2$, for the 3D wavelets set in the $k_\parallel$ vs. $k_\perp$ plane. Take the starting wavelet $\psi_{\lambda_1}$ (\textit{black}), the second wavelet $\psi_{\lambda_2}$ must probe regions such that $\mu^\parallel_{j_{\parallel,2}} \leq \mu^\parallel_{j_{\parallel,1}}$ and $\mu^\perp_{j_{\perp,2}} \leq \mu^\perp_{j_{\perp,1}}$. These wavelets, for the example $\psi_{\lambda_1}$, are shown in the highlighted region (\textit{red}).}
\label{fig:Wavelet_Scaling}
\end{figure}

Scattering transform statistics are a family of summary statistics. In this work, the scattering transform statistics we consider are a simplified version\footnote{With respect to the statistics defined in~\citep{2023:ChengMorelAllys}, no second convolution is explicitly performed. These statistics can also be seen as a compact form of the wavelet phase harmonics statistics introduced in~\citep{2020:AllysMarchandCardoso}, where no phase harmonics other than $p=0$ and $p=1$ are considered, and where no translations are performed prior to covariances estimation.} of the scattering covariances statistics introduced in~\citep{2023:ChengMorelAllys}. 
Consider a 3-dimensional lightcone $I$. Its scattering covariances statistics are defined under the unified definition:
\begin{equation}
   C^{p_1,p_2}_{\lambda_1,\lambda_2} =  \operatorname{Cov}\big( [I \circledast \psi_{\lambda_1}]^{p_1} ,[I \circledast \psi_{\lambda_2}]^{p_2} \big),
\end{equation}
where $\circledast$ denotes a convolution and,
\begin{equation}
 [z]^p =\begin{cases}
    |z|, & \text{if $p=0$}.\\
    z, & \text{if $p=1$}.
  \end{cases}\end{equation}
The covariance between two stochastic processes $\mathrm{X}$ and $\mathrm{Y}$ is defined as $\operatorname{Cov}(X, Y)=\left\langle X Y^*\right\rangle-\langle X\rangle\left\langle Y^*\right\rangle$, where $\langle \rangle$ denotes the expected value, and $^*$ denotes the complex conjugate. In this paper, expected values are estimated through spatial average.

Using this definition, the three subsets of scattering covariance coefficients are
\begin{align}
\label{EqC11}
C^{11} &= \operatorname{Cov}\big( I \circledast \psi_{\lambda_1}, I \circledast \psi_{\lambda_2} \big), \quad &\lambda_1 = \lambda_2.\\[1ex]
\label{EqC01}
C^{01} &= \operatorname{Cov}\big( |I \circledast \psi_{\lambda_1}|, I \circledast \psi_{\lambda_2} \big), \quad &\lambda_1 \leq \lambda_2,\\[1ex]
\label{EqC00}
C^{00} &= \operatorname{Cov}\big( |I \circledast \psi_{\lambda_1}|, |I \circledast \psi_{\lambda_2}| \big), \quad &\lambda_1 \leq \lambda_2,
\end{align}
In these equations, the condition $\lambda_1 \leq \lambda_2$ is a short form for $\mu^\parallel_{j_{\parallel,2}} \leq \mu^\parallel_{j_{\parallel,1}}$ and $\mu^\perp_{j_{\perp,2}} \leq \mu^\perp_{j_{\perp,1}}$, a condition which is summarised in Fig. \ref{fig:Wavelet_Scaling}. 
This conditioning on the $\lambda$ scales is explained as follows. For $C^{11}$, the covariance is zero if $\lambda_1 \neq \lambda_2$. Indeed, since the spectral supports of different wavelets do not overlap, $I \circledast \psi_{\lambda_1}$ and $I \circledast \psi_{\lambda_2}$ also have distinct spectral supports, and are thus of vanishing covariance because they are linearly decorrelated\footnote{This can also be shown using Plancherel Theorem.}. For $C^{01}$, convolution $I \circledast \psi_{\lambda_1}$ with the first wavelet concentrates the resulting field into a band of scales around $\lambda_1$. Applying the modulus operator, $|I \circledast \psi_{\lambda_1}|$, demodulates the field by re-centering its support towards lower scales, see for example~\citet{2019:ZhangMallat}. Then, in order to have a non-vanishing covariance, the secondary wavelet convolution $I \circledast \psi_{\lambda_2}$, centred at scales $\lambda_2$, must now overlap the envelope defined by $|I \circledast \psi_{\lambda_1}|$, imposing that $\lambda_1\leq\lambda_2$. For $C^{00}$, since the covariance is commutative, $|I \circledast \psi_{\lambda_1}|$ and $|I \circledast \psi_{\lambda_2}|$ play the same role, so the condition $\lambda_1 \leq \lambda_2$ is only to avoid double counting. 

%\IH{To understand the criteria that $C^{11}$ is non-zero only for $\lambda_1 = \lambda_2$, consider the Plancherel's theorem in a 1D case:
%\begin{equation}
%    \int f(x)\,g(x)^*\,dx = \int \tilde{f}(k)\,\tilde{g}(k)^*\,dk,
%\end{equation}
%where $ \tilde{f}(k) $ and $ \tilde{g}(k) $ are the Fourier transforms of $ f(x) $ and $ g(x) $, respectively. Using the Plancherel theorem, for $ C^{11} $ in 1D, we have:
%\begin{align}
%    \int (I(x) \circledast \psi_{\lambda_1}(x))\,(I(x) \circledast \psi_{\lambda_2}(x))^*\,dx 
%    &=\\ \int (\tilde{I}(k)\,\tilde{\psi}_{\lambda_1}(k))\,(\tilde{I}(k)\,\tilde{\psi}_{\lambda_2}(k))^*\,dk. \nonumber
%\end{align}
%Rewriting, we obtain:
%\begin{equation}
%    \int |\tilde{I}(k)|^2\,\tilde{\psi}_{\lambda_1}(k)\,\tilde{\psi}_{\lambda_2}^*(k)\,dk.
%\end{equation}
%The integrand contains a cross-term between the Fourier-transformed wavelets, modulated by the power spectrum of the signal $ |\tilde{I}(k)|^2 $. If $ \lambda_1 \neq \lambda_2 $, then the wavelets $ \tilde{\psi}_{\lambda_1} $ and $ \tilde{\psi}_{\lambda_2} $ are separated, and linearly decorrelated, leading to the covariance being zero.
%On the other hand, when $ \lambda_1 = \lambda_2 $, we have the non-zero:
%\begin{equation}
%    \int |\tilde{I}(k)|^2\,|\tilde{\psi}_{\lambda_1}(k)|^2\,dk.
%\end{equation}}

Although the scattering covariances $C^{p_1,p_2}$ capture the non-Gaussian properties of a given lightcone, they also depend on the amplitude of its power spectrum. To correct for this dependence, the scattering covariances are normalised by the $C^{11}$ of a reference field. In this work, this reference field will be the target lightcone, and we will call this normalised scattering covariance statistic $C^{11}_\text{ref}$. The normalised scattering covariance statistics are then defined as follows:
\begin{align}
&\bar{C}^{00}(\lambda_1,\lambda_2) = \frac{C^{00} (\lambda_1,\lambda_2)}{\sqrt{C^{11}_\text{ref}(\lambda_1)C^{11}_\text{ref}(\lambda_2)}},\\[1ex]
&\bar{C}^{01}(\lambda_1,\lambda_2) = \frac{C^{01} (\lambda_1,\lambda_2)}{\sqrt{C^{11}_\text{ref}(\lambda_1)C^{11}_\text{ref}(\lambda_2)}},\\[1ex]
&\bar{C}^{11}(\lambda) = \frac{C^{11} (\lambda)}{\sqrt{C^{11}_\text{ref}(\lambda)}}.
\end{align}

To further constrain the power spectrum, an additional set of $C^{11}$ coefficients, $C^{11 \prime}$, is used. These coefficients are computed according to Eq.~\eqref{EqC11}, with a set of wavelets defined as in Sec.~\ref{PartWavelets}, but with a different number of scales $J^\prime_\alpha > J_\alpha$. These $C^{11 \prime}$ coefficients, computed from a larger set of wavelets, allow, used in conjunction with $C^{11}$, for better overall spectral resolution for power spectrum constraints~\citep{2024:MoussetAllysPrice}, without increasing the computational complexity for higher-order terms. For these coefficients, we used $(J^{\prime}_\parallel,J^{\prime}_\perp) = (15,15)$, corresponding to $(Q_\parallel,Q_\perp) = (2.53,2.60)$. -

In the following, we concatenate the various ST coefficients introduced below to form the set of ST statistics:
\begin{equation}
    \phi(I) = \Bigl\{\bar{C}^{11},\bar{C}^{11\prime},\bar{C}^{00},\bar{C}^{01} \Bigl\}.
\end{equation}
The choices of parameters to define the wavelet sets then result in 36 coefficients in \({C}^{11}\), 225 coefficients in \({C}^{11\prime}\), and 441 coefficients in \({C}^{01}\) and \({C}^{00}\), for a total of 1143 ST statistics.

%%%%%%%%%%%%%%%%%%%%%%%%%%%%%%%%%%%%%%%%%%%%%%%%%%%%%%%%%%%%%%%%%%%%%%%%%%%%%%
%%%%%%%%%%%%%%%%%%%%%%%%%%%%%%%%%%%%%%%%%%%%%%%%%%%%%%%%%%%%%%%%%%%%%%%%%%%%%%
\section{Generative Models}
\label{Method}
%%%%%%%%%%%%%%%%%%%%%%%%%%%%%%%%%%%%%%%%%%%%%%%%%%%%%%%%%%%%%%%%%%%%%%%%%%%%%%
\subsection{Maximum Entropy Microcanonical Models}

In this paper, we construct a maximum entropy model of EoR lightcone from the scattering covariance statistics on the target lightcone $d_t$. New approximate realisations of this lightcone are then sampled by gradient descent~\citep{2018:BrunaMallat,2020:AllysMarchandCardoso,2023:BrunoAllysAuclair}. This gradient descent sampling consists in transforming a white Gaussian distribution into the maximum entropy distribution conditioned on the ST statistics of $d_t$. 

The procedure is as follows. The map on which the gradient descent is performed is called $u$. It is initialised with a white noise realisation, $u=u_0$, with a mean and standard deviation matching those of the target lightcone. The optimisation is then performed by constraining the scattering transform statistic $\phi(u)$ of $u$ with the following loss:
\begin{equation}
\label{EqLoss}
    \mathcal{L}(u)=\|\phi(d_t)-\phi(u)\|^2,
\end{equation}
where $\|\cdot\|$ is the Euclidean norm, where $d_t$ is fixed. This gradient descent is performed until the loss converges. The samples $\tilde{d}$ from the ST generative model correspond to the map $\tilde{d}= u_f$ obtained at the end of this optimisation, whose ST statistics are close to those of $d_t$. Different $\tilde{d}$  samples are obtained by starting the optimisation with different white noise realisations.

%%%%%%%%%%%%%%%%%%%%%%%%%%%%%%%%%%%%%%%%%%%%%%%%%%%%%%%%%%%%%%%%%%%%%%%%%%%%%%
\subsection{Data Conditioning}

Prior to the construction of the generative model, the data is conditioned to mitigate several practical issues that can adversely affect the resulting synthesis. This data conditioning ensures that the generative model operates on a well-suited representation of the EoR lightcone, thus improving the fidelity and stability of the subsequent synthesis.

%%%%%%%%%%%%%%%%%%%%%%%%%%%%%%%%%%%%%%%%%%%%%%%%%%%%%%%%%%%%%%%%%%%%%%%%%%%%%%
\subsubsection{Quantile Function}

The EoR lightcone is a complex non-Gaussian process, in particular with a highly-skewed histogram, the latter property being difficult to recover with a ST maximum entropy model. To improve the modelling, we first transform the EoR lightcone, using an invertible non-linear transform, to obtain a histogram closer to a Gaussian distribution. We then sample the ST maximum entropy model in this transformed data space, before returning to the original EoR lightcone data space using the inverse non-linear transform.

Such ST modelling based on an invertible non-linear transform has been used in previous work, for example using a logarithmic pointwise transform before constructing ST generative models, as in~\cite{2020:AllysMarchandCardoso,2023:BrunoAllysAuclair,2024:MoussetAllysPrice}. It is particularly useful for modelling processes with complex histograms, such as those with only non-negative values, like the matter density of the Large Scale Structures of the Universe. We note that this idea of transforming the process under study into another one that can be better approximated is also used when constructing lognormal models, for processes whose logarithm is expected to be much better described by a Gaussian model than the process itself. Lognormal models are used, for example, for describing the thermal emission of Galactic molecular clouds~\citep{miville2007statistical,levrier_statistics_2018}.

In this paper, we use a quantile transformation to map the histogram of the target lightcone to a normal distribution~\citep{2009:Beasley,2020:McCullaghTresoldi}. To do this, the global empirical Cumulative Distribution Function (CDF) \(\hat{F}\) of the field is first estimated for the entire ($\delta \nu \sim 14 MHz$) lightcone . Then, for each data point \(d(\vec{r})\) in the lightcone, the corresponding percentile \(p\) is found:
\begin{equation} 
p(\vec{r}) = \hat{F}\left(d(\vec{r})\right).
\end{equation}
Next, this percentile \(p\) is passed into the inverse of the standard normal CDF (also known as the quantile function) to obtain:
\begin{equation}
\bar{d}(\vec{r}) = \Phi^{-1}(p(\vec{r})).
\end{equation}
In this way, the resulting set of \(\bar{d}\) values matches a normal distribution, replacing the originally skewed one, while preserving the ordering of the original data points. 

In this work, the quantiles used in the quantile transform are estimated from all data points within the target lightcone. The choice of the number of quantiles corresponds to a usual bias/variance trade-off: too few quantiles result in a poor representation of the data distribution, while too many can introduce overfitting and unnecessary computational overhead. We empirically find that 10,000 quantiles provide a good balance. Given that the target lightcone contains approximately $10^{6}$ pixels, using 10,000 quantiles corresponds to approximately 300 pixels per quantile. This quantile to pixel ratio is sufficient to accurately represent the low density tail of the distribution without incurring needless computational cost.

It should be noted that the quantile transform is not a pointwise operator, but is instead applied by analysing the entire lightcone simultaneously. Although this transform is very efficient for modelling complex histograms, this means that any modelling error in the transformed data space can lead to significant instabilities in the inverse transform. In particular, the significant effect of large scales on the histogram, which are difficult to model statistically from a small amount of data, can make it difficult to use such a transform.

\begin{figure*}[!ht]
  \centering
  \textbf{\large XY-plane}

  \begin{subfigure}[b]{0.24\textwidth}
    \centering
    \caption{\fontsize{11}{11}\selectfont Target Field}
    \includegraphics[width=\linewidth]{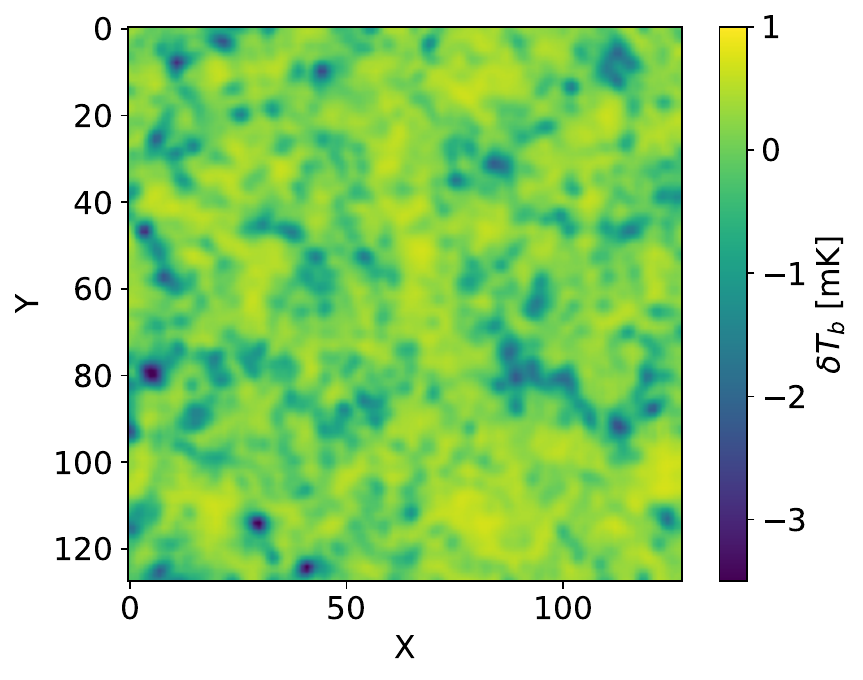}
  \end{subfigure}\hfill
  \begin{subfigure}[b]{0.24\textwidth}
    \centering
    \caption{\fontsize{11}{11}\selectfont Synthesised Field 1}
    \includegraphics[width=\linewidth]{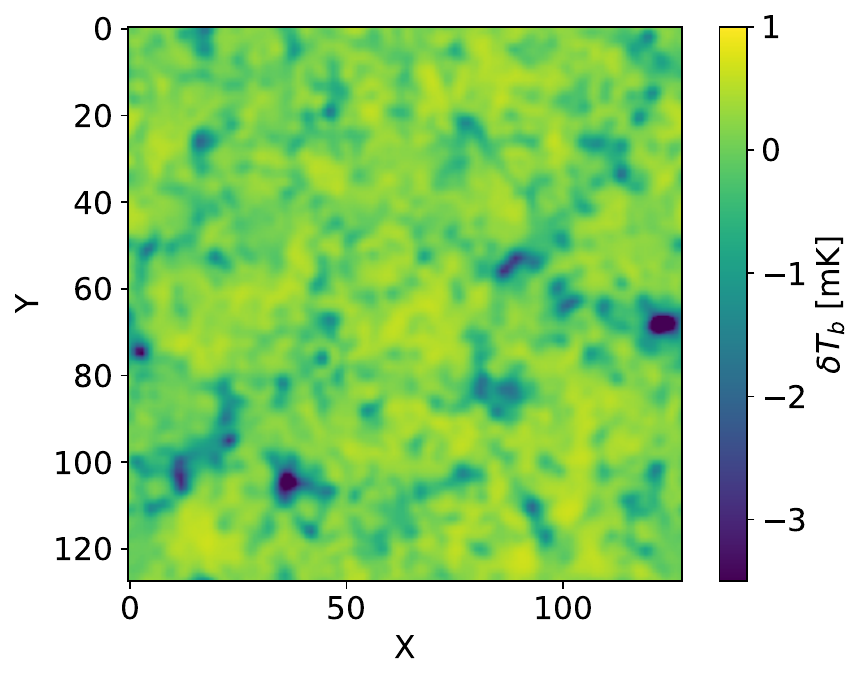}
  \end{subfigure}\hfill
  \begin{subfigure}[b]{0.24\textwidth}
    \centering
    \caption{\fontsize{11}{11}\selectfont Synthesised Field 2}
    \includegraphics[width=\linewidth]{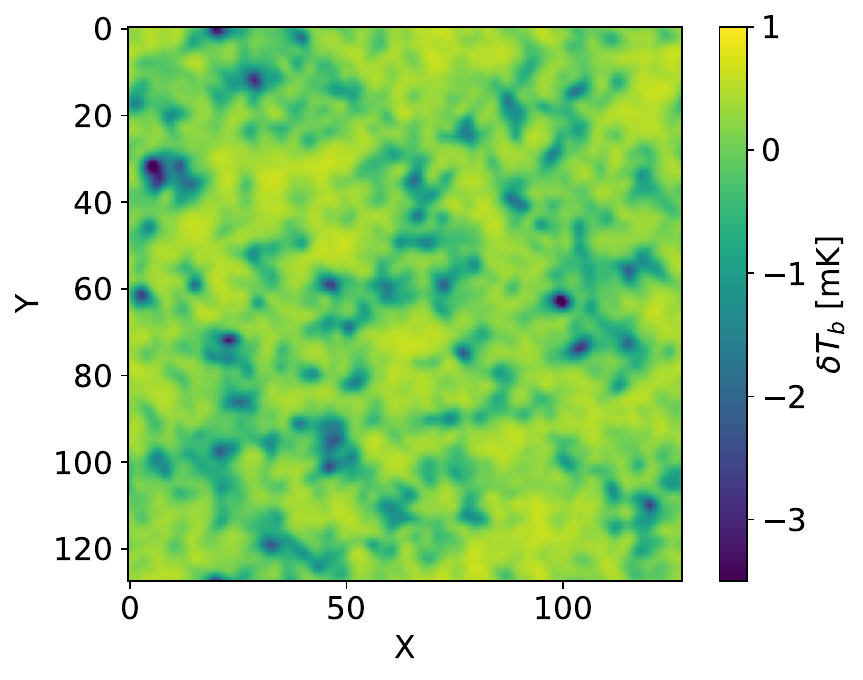}
  \end{subfigure}\hfill
  \begin{subfigure}[b]{0.24\textwidth}
    \centering
    \caption{\fontsize{11}{11}\selectfont Synthesised Field 3}
    \includegraphics[width=\linewidth]{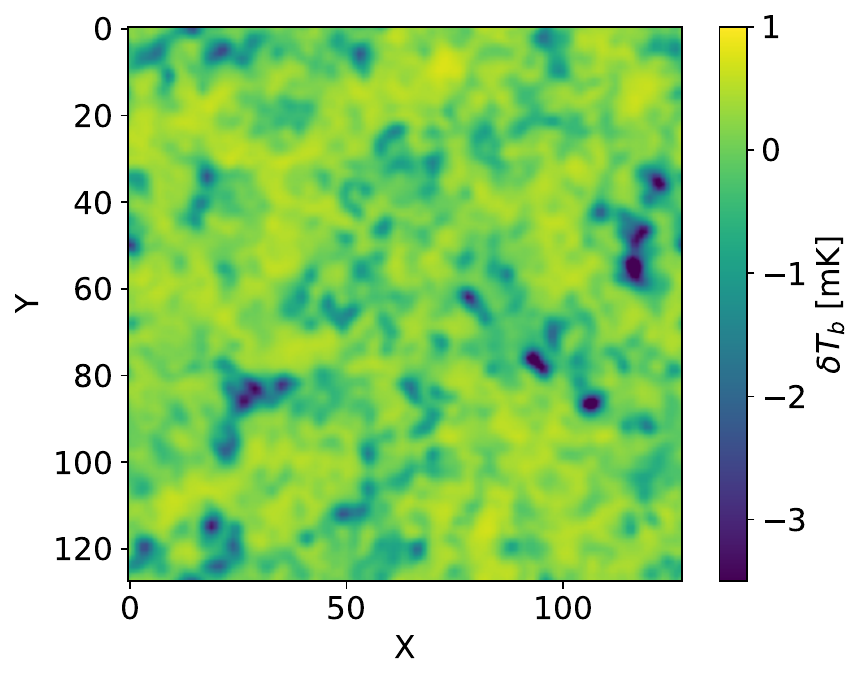}
  \end{subfigure}

  \caption{Comparison of a given $xy$-slice (single frequency channel in the lightcone), for the target EoR lightcone and three lightcones that have been synthesised using the 3D scattering transforms and have had the inverse quantile transform applied to them. The synthesised lightcones (visually) reproduces target lightcone well.}
  \label{fig:LoReLi_Comparison_XY_Lightcone}
\end{figure*}

\begin{figure*}[!ht]% or \begin{figure} if you don't need full‐page width
  \centering
  \textbf{\large XZ-plane}

  \begin{subfigure}[b]{0.24\textwidth}
    \centering
    \caption{\fontsize{11}{11}\selectfont Target Field}
    \includegraphics[width=\linewidth]{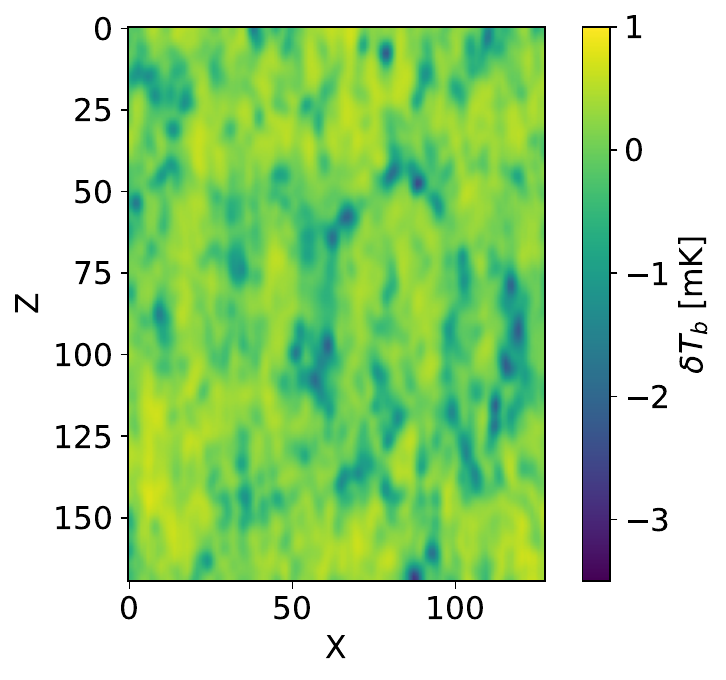}
  \end{subfigure}\hfill
  \begin{subfigure}[b]{0.24\textwidth}
    \centering
    \caption{\fontsize{11}{11}\selectfont Synthesised Field 1}
    \includegraphics[width=\linewidth]{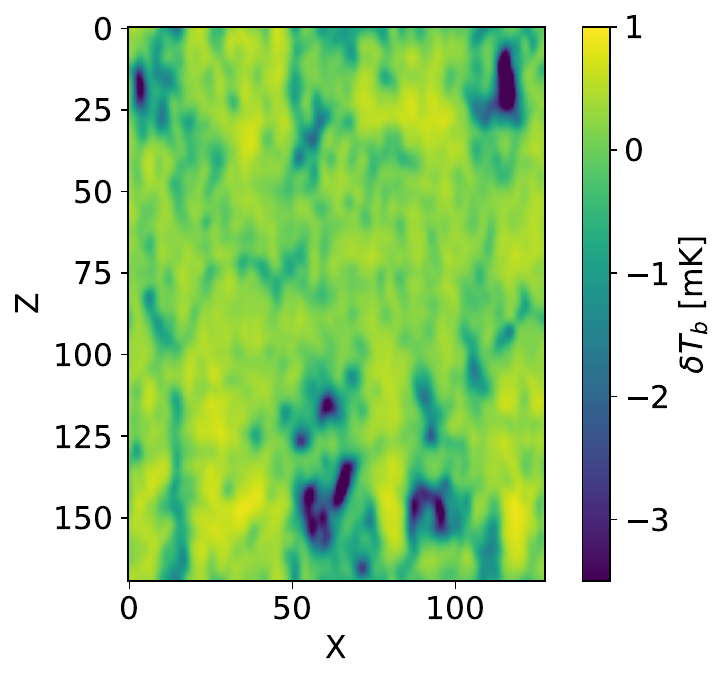}
  \end{subfigure}\hfill
  \begin{subfigure}[b]{0.24\textwidth}
    \centering
    \caption{\fontsize{11}{11}\selectfont Synthesised Field 2}
    \includegraphics[width=\linewidth]{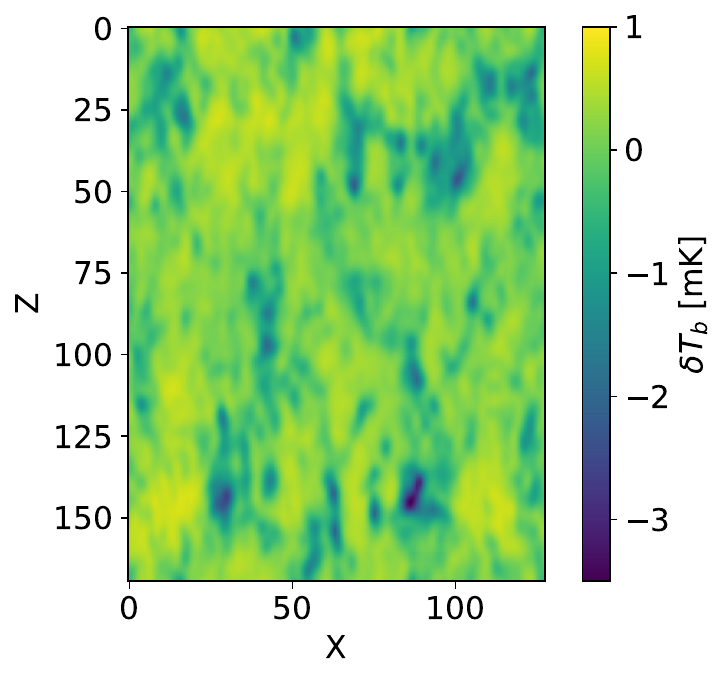}
  \end{subfigure}\hfill
  \begin{subfigure}[b]{0.24\textwidth}
    \centering
    \caption{\fontsize{11}{11}\selectfont Synthesised Field 3}
    \includegraphics[width=\linewidth]{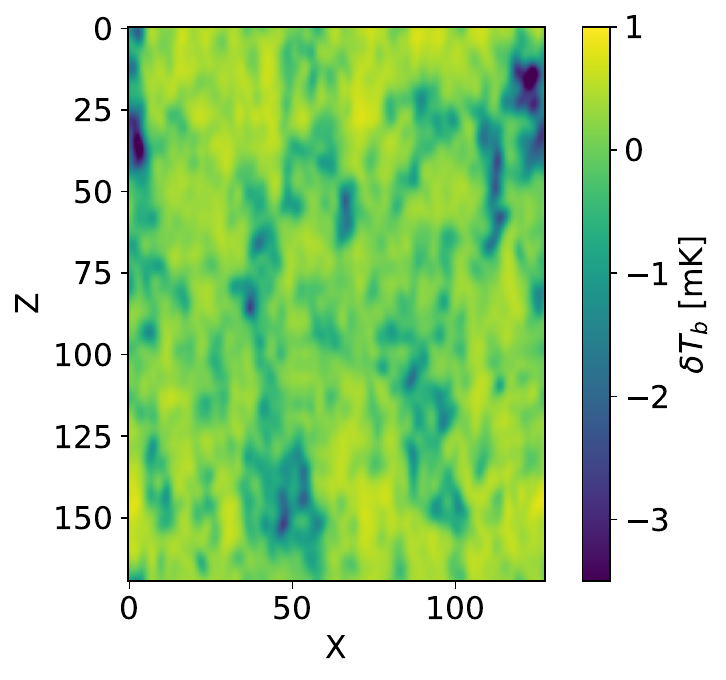}
  \end{subfigure}

  \caption{Same as Figure \ref{fig:LoReLi_Comparison_XY_Lightcone} but for a given $xz$-slice (the $z$-domain being the redshift-domain), for the target EoR lightcone and three synthesised lightcones. As with \ref{fig:LoReLi_Comparison_XY_Lightcone}, the target lightcone is visually well produced by the synthesised lightcones.}
  \label{fig:LoReLi_Comparison_XZ_Lightcone}
\end{figure*}

%%%%%%%%%%%%%%%%%%%%%%%%%%%%%%%%%%%%%%%%%%%%%%%%%%%%%%%%%%%%%%%%%%%%%%%%%%%%%%
\subsubsection{Treatment of unconstrained Scales}
\label{limiting_scales}
There are two limiting ranges of scales in the data that the generative model of this paper does not constrain. The first range of unconstrained scales, denoted $\kappa_{low}$, corresponds to largest scales, which have Fourier modes wavenumbers smaller than the mother wavelet's wavenumber $\mu_0$. The second range of unconstrained scales, denoted $\kappa_{high}$, corresponds to smallest scales, which have Fourier modes beyond the Nyquist frequency. 
%Consider the 2D sky-plane in the 3D field, defined along the $x$- and $y$-axes. When Fourier transforming this plane, the maximum wavenumbers $k_{x,\text{max}}$ and $k_{y,\text{max}}$ are set by the resolution of the field. In two-dimensional Fourier space, however, there exist modes with wavenumber larger than $\sqrt{k_{x,\text{max}}^2 + k_{y,\text{max}}^2}$ and these scales exceed the Nyquist frequency. 
\begin{comment}
Although $k_{x,\text{max}}$ and $k_{y,\text{max}}$ themselves lie within the Nyquist limit, any mode beyond $\sqrt{k_{x,\text{max}}^2 + k_{y,\text{max}}^2}$ is unconstrained by the data. 
\end{comment}

In this paper, as we chose to first construct the ST generative model on the quantile transformed data, before applying the inverse quantile transform, these unconstrained scales can significantly affect the final synthesised lightcone, owing to the non-linear and non-local nature of this transform. To mitigate the impact of these ranges of scales, we handle them explicitly, as described below:
\begin{itemize}
    %\item The $\kappa_{low}$ scales are replaced by the corresponding modes from the target lightcone in the transformed data space. This is because these scales, if not constrained, will affect the global inverse quantile transform as described above. However, as these scales are not characterised by ST statistics, they have no direct effect on how the other scales are generated.

    \item The $\kappa_{low}$ scales are replaced by the corresponding modes from the target lightcone in the transformed data space. This substitution occurs prior to the gradient descent optimisation, ensuring that these large scale modes are included in the initial conditions. This is necessary because, if left unconstrained, these scales would influence the global inverse quantile transform as described previously. However, as these scales are not characterised by the ST statistics, they do not directly influence the generation of the remaining scales during the optimisation process.

    \item The $\kappa_{high}$ scales in the target lightcone are filtered out before the quantile transform, both when constructing the generative model and for the comparison with the synthesised maps. Due to the non-linearity of the quantile transform, there will still be some beyond-Nyquist scales after the histogram transform, but their effect on the inverse histogram transform after modelling is negligible.
\end{itemize}

%%%%%%%%%%%%%%%%%%%%%%%%%%%%%%%%%%%%%%%%%%%%%%%%%%%%%%%%%%%%%%%%%%%%%%%%%%%%%%
\subsection{Non-periodic boundary conditions.}
\label{Part_NonPBC}

Simulations of the EoR lightcone have non-periodic boundaries, in particular due to the redshift evolution along the line of sight.
However, since we compute convolutions by products in Fourier space, these are inherently periodic. To avoid taking boundary discontinuities into account when estimating the ST statistics, the empirical covariances are computed only over a central smaller cubic volume of the data\footnote{In practice a rectangular parallelepiped, as the original lightcone.}. This is done by cropping the edges of the wavelet convolved fields, so that all pixels over which the average is computed are sufficiently separated from the boundaries. To verify this, the cropped band sizes are matched to the maximum sizes characterised by the wavelets, which are determined by the mother wavelet scale $\frac{1}{\mu_0}$. Specifically, we remove a width at each end of each axis equal to half the mother wavelet scale on that axis, leading to an average over a volume of (61,61,87) pixels.

%\IH{Since convolutions are periodic, wavelet convolutions performed during scattering transformations wrap around the edges of the non-periodic field, causing small-scale artefacts in the synthesised lightcone. To mitigate this effect, we crop the field size along each axis by an amount equal to the largest wavelet scale sampled, as determined by the mother wavelet scale $\frac{1}{\mu_0}$. Specifically, we remove an equal width at each end of each axis, where the width is half the mother wavelet scale, moving all non-zero data away from the boundaries. Performing the convolutions on this reduced field minimises the wrap-around artefacts at these scales, as the convolution no longer has non-zero values at the edges to propagate through the periodicity.} 

Note that while the statistics of the target lightcones, estimated on $d_t$ and used to constrain the generative model, are computed taking into account the non-periodic boundary conditions as described above, the samples generated by this generative model are full lightcones with periodic boundary conditions. To do this, their initial conditions are white noise defined over the entire (128,128,170) pixels data cube, and their statistics are computed using period convolutions and from averages over the entire cube. These syntheses are performed by comparing the two types of ST statistics estimators, with and without periodic boundary conditions, in Eq.~\eqref{EqLoss}. As a side effect, we note that the statistics on which the generative models are constructed are estimated on only about one eighth of the total volume of the target light cone, leading to a higher sampling variance.

\begin{figure*}
  \centering
  \begin{subfigure}[b]{0.49\textwidth}
    \centering
    \caption{Linear Scale}
    \includegraphics[width=\linewidth,height=8cm,keepaspectratio]{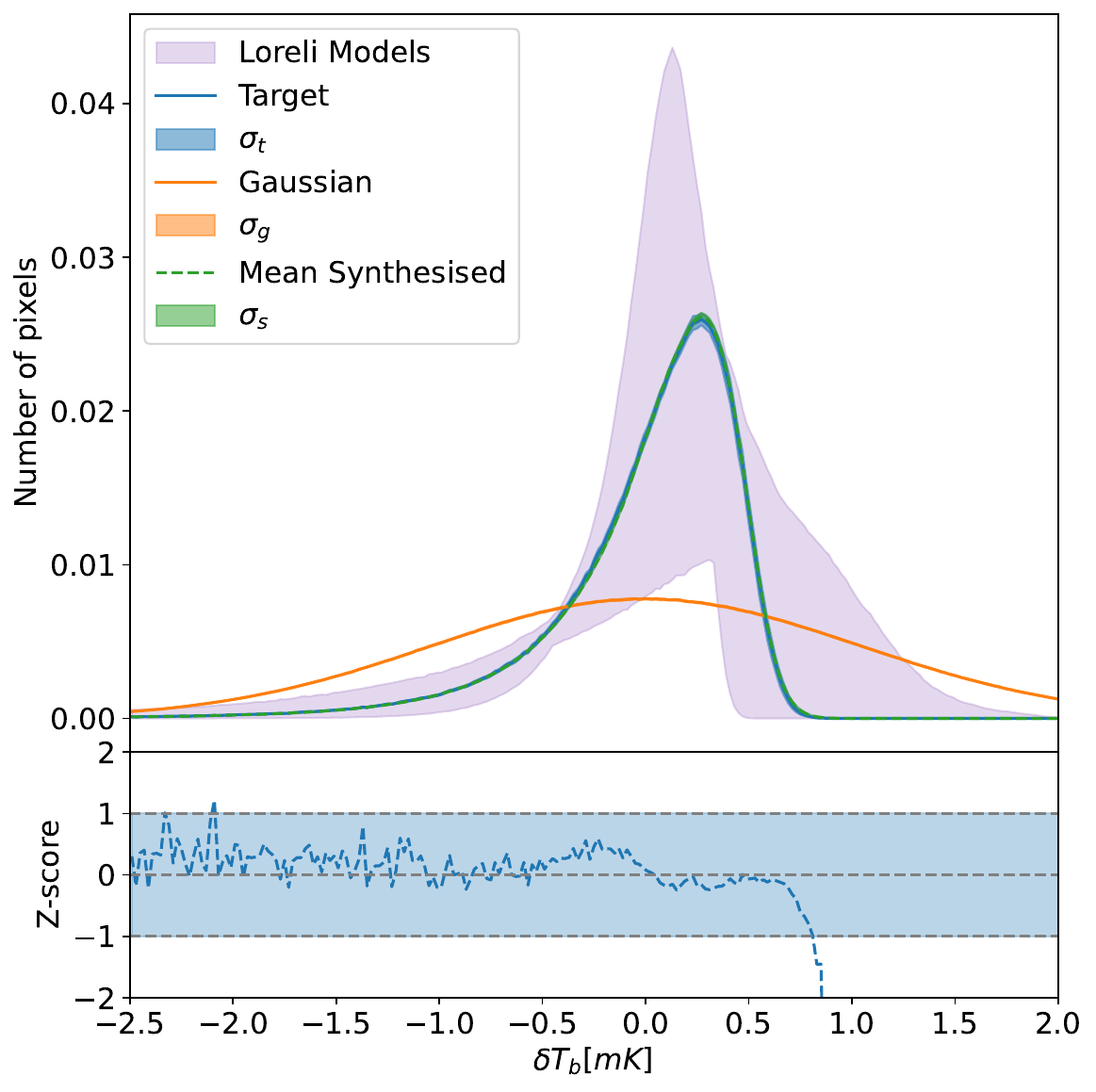}
    \label{fig:linear}
  \end{subfigure}
  \begin{subfigure}[b]{0.49\textwidth}
    \centering
    \caption{Log Scale}
    \includegraphics[width=\linewidth,height=8cm,keepaspectratio]{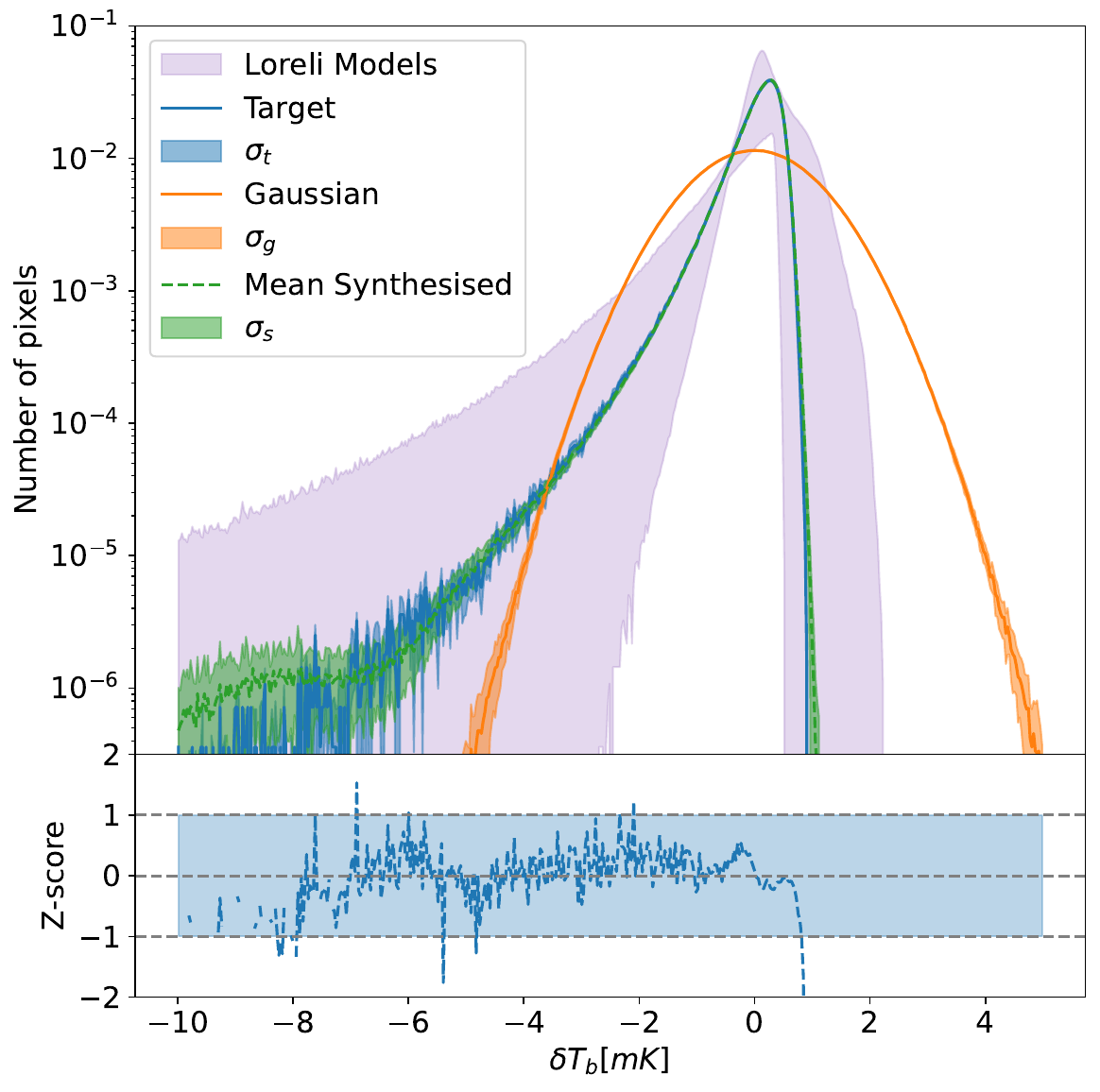}
    \label{fig:log}
  \end{subfigure}\hfill

  \caption{The linear-scale and log-scale histograms of the target EoR lightcone and synthesised lightcone, post-inverse quantile transform. The histogram of the synthesised lightcone, for which the mean of 30 realisations and their standard deviation (shaded region) are shown, can recover the histogram of the target lightcone well, including the target lightcones’ skewness.}
  \label{fig:Synth_Histo}
\end{figure*}

%%%%%%%%%%%%%%%%%%%%%%%%%%%%%%%%%%%%%%%%%%%%%%%%%%%%%%%%%%%%%%%%%%%%%%%%%%%%%%
%%%%%%%%%%%%%%%%%%%%%%%%%%%%%%%%%%%%%%%%%%%%%%%%%%%%%%%%%%%%%%%%%%%%%%%%%%%%%%
\section{Validation of the generative models}
\label{Results}

\subsection{Numerical experiment and statistical validation}

In this section, we validate the quality of the ST generative models by comparing the statistical properties of the generated lightcones, after performing the inverse quantile transform, to those of the target lightcone $d_t$. To do so, we generate 30 independent synthesised lightcones, noted $\left\{\tilde{d}_i\right\}_{i=1\cdots 30}$ after the inverse quantile transform, each starting from a different white Gaussian noise realisation. To perform the optimisation, the gradient of the loss function is computed by automatic differentiation using the L-BFGS-B algorithm~\citep{LBFGS} (as implemented in PyTorch). Optimisation is done by performing 2000 gradient descent iterations, at which point the loss has converged to a plateau after the loss has been reduced by $\sim 10$ orders of magnitude. Each such optimisation for a single lightcone takes 2 hours on an Nvidia Tesla A100 GPU with 80 GB of memory.

Beyond a first visual comparison, the statistics to be compared are histogram, power spectrum and Minkowski Functionals. In each case, the mean statistics are estimated on $d_t$ (in blue) and on the 30 $\tilde{d}_i$ (in green) are compared. To assess the presence of any model bias, the difference between these mean statistics is compared, where possible, with their sample standard deviation $\sigma_t^\text{oct}$ from one octant of the lightcone to another\footnote{To create octants of the light cone, the data cube is divided into eight equal sub-volumes, of dimension (64,64,85) pixels. These octants have approximately the same volume as the one on which the ST statistics are estimated due to the non-periodic boundary conditions.},
%As the lightcone is represented as a three-dimensional cube, this is achieved by splitting each of the three spatial axes at their mid-points. The result is eight subcubes, each corresponding to an octant, with dimensions half the size of the full lightcone along each axis, and each octant occupying an eighth of the total volume of the lightcone.}}
estimated on the target lightcone $d_t$. Indeed, the standard deviation of the estimator of the ST statistics used to construct the generative model is $\sigma_t^\text{oct}$, since only a volume corresponding to a single octant of the lightcone is used, due to the issue of non-periodic boundary conditions, as explained above. This sampling variance could then simply explain a modelling bias of this order.

In addition, a comparison will be made between $\sigma_t$ and $\sigma_s$, the sample standard deviation from one lightcone to another, estimated from $d_t$ and the 30 $\tilde{d}_i$, respectively. This will allow to quantify whether the ST generative model reproduces the sample variance of the statistics listed below, in addition to their mean value. While $\sigma_s$ can be estimated directly from the 30 $\tilde{d}_i$, $\sigma_t$ has to be estimated from $d_t$ only. To do this, we make the rough approximation that the octants of $d_t$ are independent, and thus consider that $\sigma_t \simeq \sigma_t^\text{oct} / \sqrt{8}$. While this is a rough approximation, which will a priori tend to underestimate $\sigma_t$, since the octant are not independent, we still believe that it allows for a first comparison between $\sigma_t$ and $\sigma_s$.

\begin{table}[h]
  \caption{Overview of the $\sigma$ quantities defined}
  \label{tab:sigma_defs}
  \centering
  \small
  \begin{tabular}{p{0.22\columnwidth} p{0.68\columnwidth}}
    \hline\hline
    Symbol & Definition \\
    \hline
    $\sigma_{\mathrm{oct},t}$ 
      & Sample standard deviation across the eight octants of the target lightcone ${d}_t$. \\[1ex]

    $\sigma_{t}$ 
      & Approximate lightcone‐to‐lightcone standard deviation for $\mathbf{d}_t$, given by 
      $\displaystyle \sigma_{t} \approx \sigma_{\mathrm{oct},t}/\sqrt{8}$. \\[1ex]

    $\sigma_{s}$ 
      & Lightcone‐to‐lightcone standard deviation from 30 independent synthesised lightcones 
      $\{\tilde{\mathbf{d}}_{i}\}$. \\[1ex]

    $\sigma_{g}$ 
      & Lightcone‐to‐lightcone standard deviation from 30 Gaussian realisations, having the same power spectrum as ${d}_t$. \\
    \hline
  \end{tabular}
\end{table}

For a fuller discussion, comparisons are also made with the distributions of statistics obtained from 30 Gaussian realisations with the same power spectrum as $d_t$ (shown in orange), with their associated lightcone-to-lightcone sample standard deviation $\sigma_g$, as well as with the distributions obtained from the full set of light cones in the Loreli II database (shown in purple).
\newline
A list of the different $\sigma$ values defined in this section can be found in Table \ref{tab:sigma_defs}.
\subsection{Visual validation}

One of the first tests of the quality of the synthesis is to visually compare three realisations $\tilde{d}_i$ of a synthesised lightcone with the target lightcone $d_t$. The comparison of the $xy$-slices, Fig. \ref{fig:LoReLi_Comparison_XY_Lightcone}, shows that these three realisations look visually similar to $d_t$. Indeed, the target lightcone contains several distinctive features, such as strongly localised absorbing (negative) regions and ionised (near-zero) regions, which can be used as touchstones that the synthesised lightcones successfully capture. Comparing the $xz$-slices, Fig. \ref{fig:LoReLi_Comparison_XZ_Lightcone}, the synthesised lightcones again contain the different phases of the field (emission, absorption, ionised) as well as the elongated structures characteristic of the redshift axis. However, it should be noted that the generated samples seem to show a greater diversity in the distribution of ionised regions, although it is difficult to make a quantitative claim from a single $d_t$ lightcone.

%%%%%%%%%%%%%%%%%%%%%%%%%%%%%%%%%%%%%%%%%%%%%%%%%%%%%%%%%%%%%%%%%%%%%%%%%%%%%%

\subsubsection{Histogram}

Fig.~\ref{fig:Synth_Histo} shows the comparison of normalised histograms, (i.e., the estimated probability density functions) of the target (blue) and synthesised (green) light cones. The target lightcone's non-symmetric, left-skewed tail distribution is a clear, highly non-Gaussian signature, which is accurately reproduced by the synthesised ones over about four orders of magnitude, down to $\delta T_b = -7$ mK. Indeed, the bottom plot of Fig.~\ref{fig:Synth_Histo} shows that the difference between these histograms, normalised by $\sigma_t^\text{oct}$, mostly lies between -1 and 1, which is compatible with the estimation variance of the target ST statistics. The top plot of Fig.~\ref{fig:Synth_Histo} also shows that $\sigma_s$, the lightcone-to-lightcone standard deviation for the synthesised maps, is of the same order as the bin-to-bin variability of the histogram estimated on $d_t$, indicating that this variance is also well reproduced. We chose not to indicate the standard deviation $\sigma_t$ in the top plot, since the sample variability is already visible due to the fine binning choice.

%Furthermore, the mean histogram of the synthesised lightcones closely approximates the target histogram, with a deviation of less than 20% for values of $\delta T_b \leq 0.5$ mK.

%The ratio subplots further confirm that the mean histogram of the synthesised data is consistently within the standard deviation (represented by the blue shaded region) of the target histogram. This indicates that the histogram deviation of the generative model is within the expected sample variance of the target histograms.

\begin{comment}  
For values 0.5 mK < $\delta T_b$, the mean histogram diverges from the target, but the variance of the synthesised lightcone histograms also increases, with the limit of $+1\sigma$ close to ratio values of 0.75. From the histogram comparison in log scale, shown in Fig. \ref{fig:log}, the mean histogram of the synthesised lightcones can reproduce the target lightcones histogram within 20$\%$ between $-5 ~\text{mK} \leq \delta T_b \leq 1~ \text{mK}$. The tail of the mean histogram, $\delta T_b < -5 ~\text{mK}$, has a marked increase in variance redolent of the effect of sampling variance on the target histogram at these values of $\delta T_b$. 
\end{comment}

%%%%%%%%%%%%%%%%%%%%%%%%%%%%%%%%%%%%%%%%%%%%%%%%%%%%%%%%%%%%%%%%%%%%%%%%%%%%%%

\subsubsection{Power Spectrum}

\begin{figure*}[!htb]
  \centering
  % Subfigure (a)
  \begin{subfigure}[b]{0.32\textwidth}
    \centering
    \includegraphics[width=\linewidth,height=7cm]{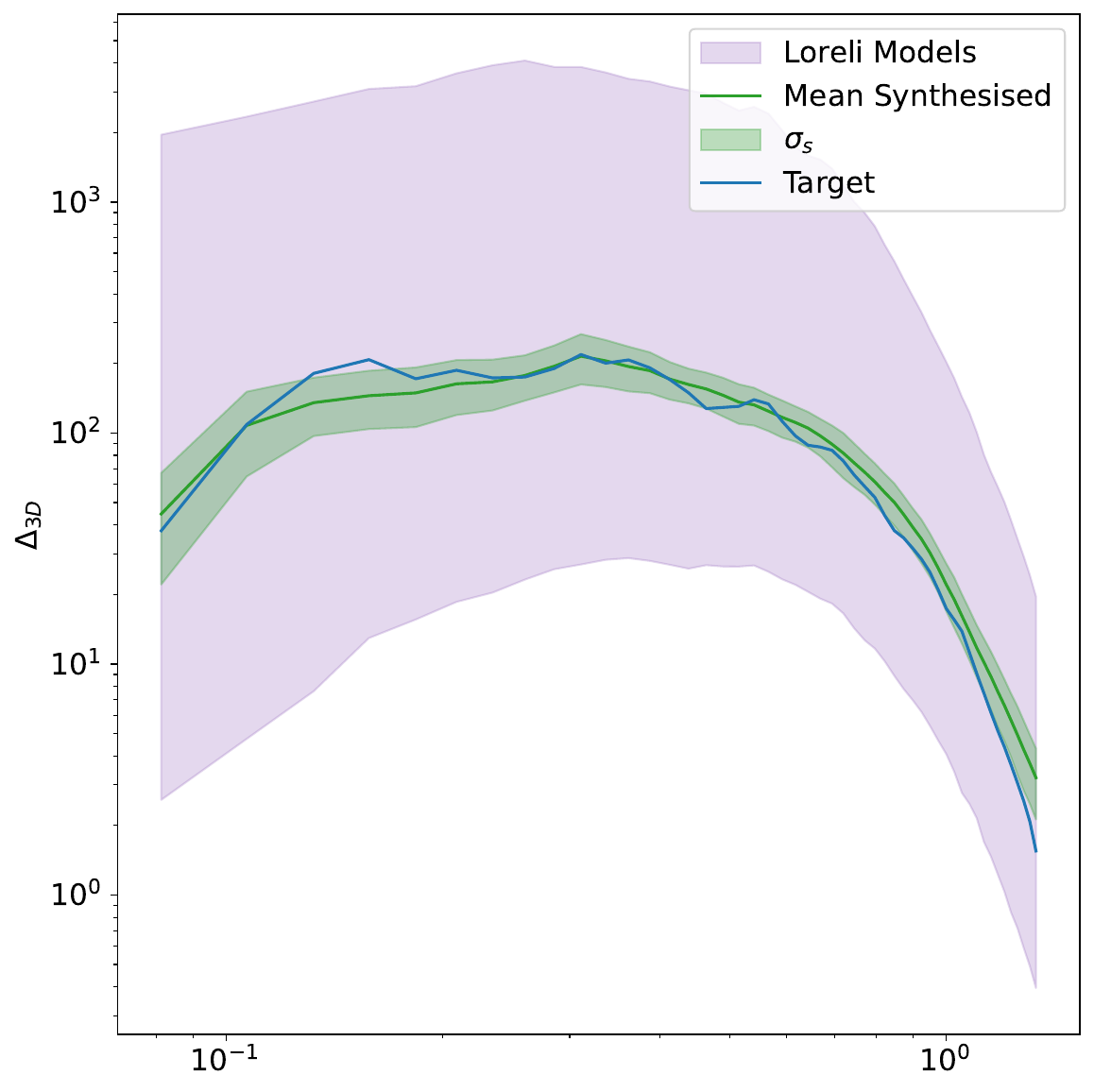}
    \caption{3D power spectrum.}
    \label{fig:PS_3D}
  \end{subfigure}\hfill
  % Subfigure (b)
  \begin{subfigure}[b]{0.32\textwidth}
    \centering
    \includegraphics[width=\linewidth,height=7cm]{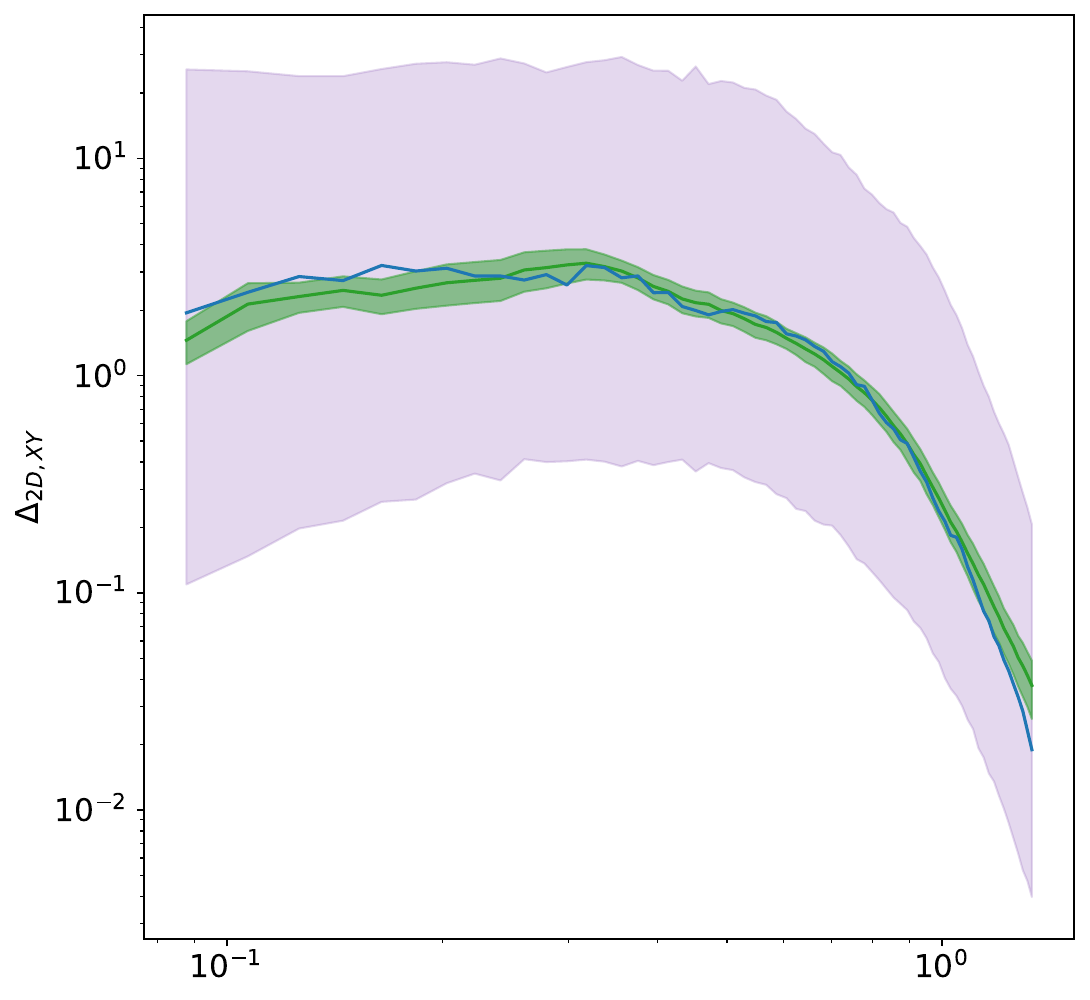}
    
    \caption{Mean 2D power spectrum on $xy$‐planes.}
    \label{fig:PS_2D_XY_Comp}
  \end{subfigure}\hfill
  % Subfigure (c)
  \begin{subfigure}[b]{0.32\textwidth}
    \centering
    \includegraphics[width=\linewidth,height=7cm]{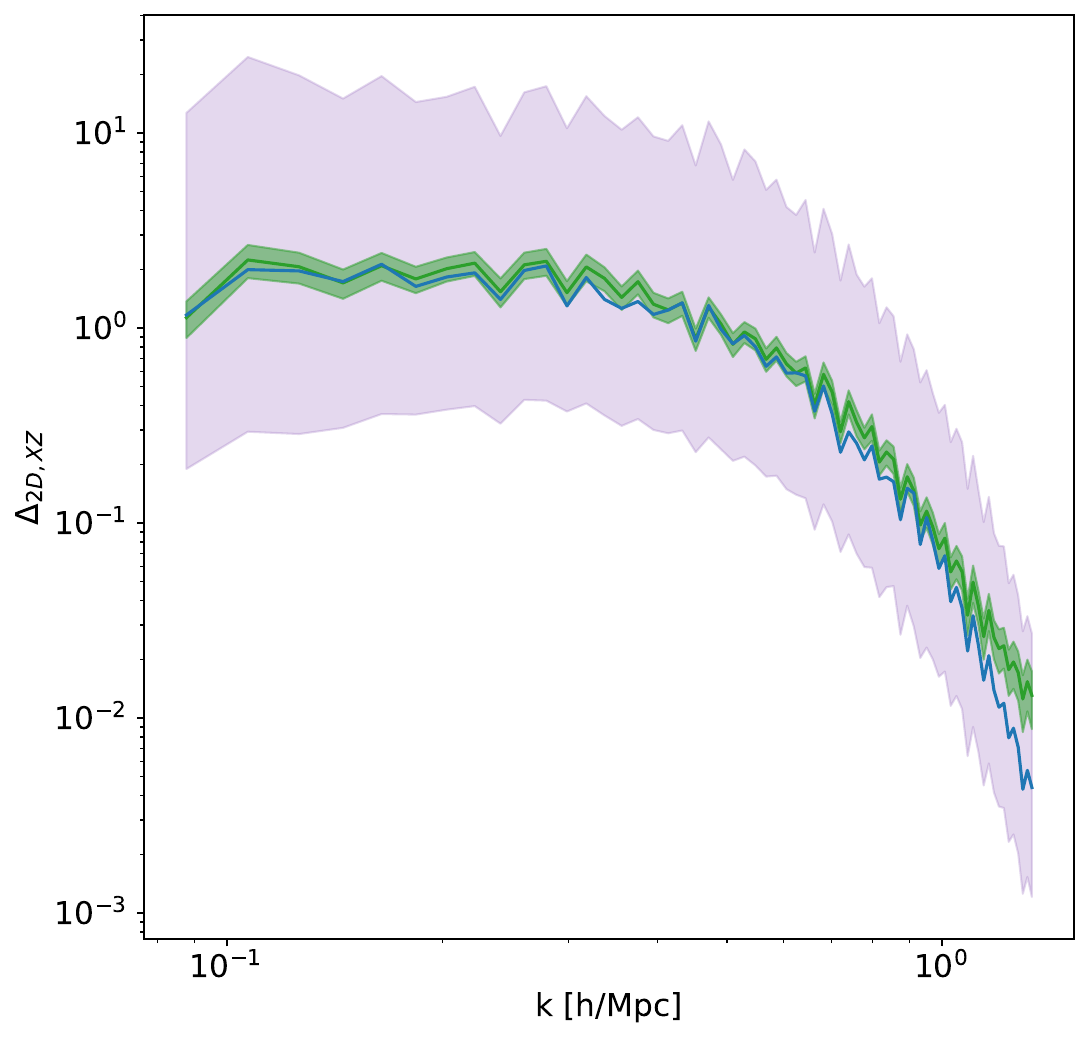}
    \caption{Mean 2D power spectrum on $xz$‐planes.}
    \label{fig:PS_2D_XZ_Comp}
  \end{subfigure}

  \caption{Comparison between the target and synthesised lightcones, post-inverse quantile transform, power spectra across different dimensions. (a) Spherically‐averaged 3D power spectrum. (b) 2D power spectrum estimated on $xy$‐planes. (c) 2D power spectrum estimated on $xz$‐planes. In all cases the synthesised lightcones power spectrum reproduces the largest scales well, with increasing deviations at smaller scales.}
  \label{fig:PS_combined}
\end{figure*}

\begin{figure*}
\centering
\begin{subfigure}{.33\textwidth}
  \centering
  \caption{Target}
  \includegraphics[width=1.1\linewidth]{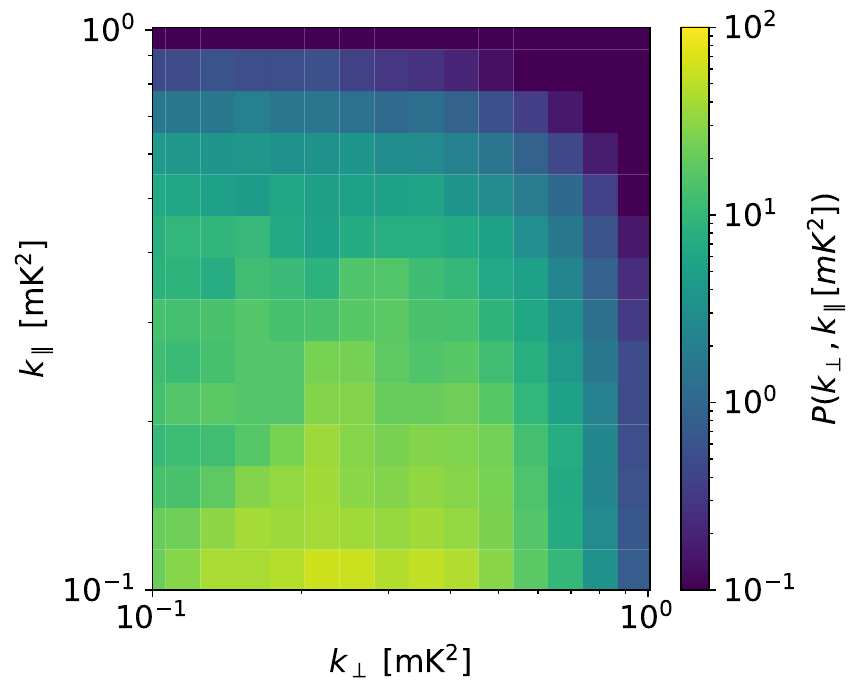}
  
\end{subfigure}
\begin{subfigure}{.33\textwidth}
  \centering
  \caption{Synthesised}
  \includegraphics[width=1.1\linewidth]{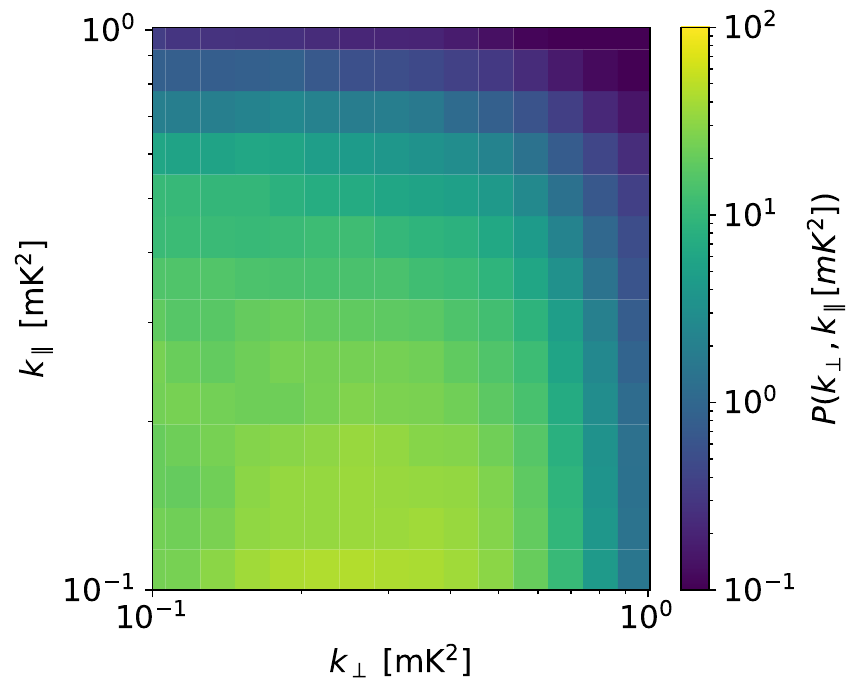}
  
\end{subfigure}
\begin{subfigure}{.33\textwidth}
  \centering
   \caption{Ratio}
  \includegraphics[width=1.1\linewidth]{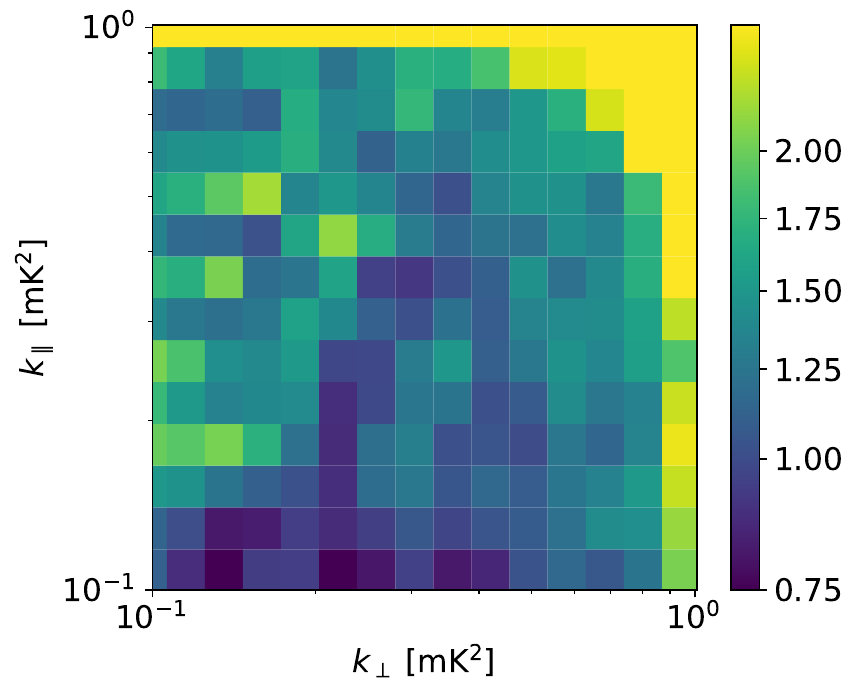}
  \label{ratio}
 
\end{subfigure}
\caption{The comparison of the cylindrically-averaged power spectrum of the target lightcone (\textit{left}) and synthesised lightcones (\textit{middle}), post-inverse quantile transform. The ratio between the two cylindrically-averaged power spectra are also presented (\textit{right}), with the synthesised lightcones power spectra reproducing that of the target lightcone within 25$\%$ on most scales below $k_\parallel = 0.3$ $h$Mpc$^{-1}$.}
\label{fig:Cylin_PS_Comp}
\end{figure*}

Four power spectra are used for comparing the target and synthesised lightcones: the three-dimensional (3D) isotropic power spectrum; the two-dimensional (2D) isotropic power spectrum on the \(xy\) or \(xz\) planes; and the cylindrically averaged power spectrum. Although the cylindrically averaged power spectrum is sometimes referred to as the 2D power spectrum, in this paper it refers to the power spectrum estimated on actual 2D fields: for each 2D \(xy\) or \(xz\) slice, and then averaged over the \(z\) or \(y\) axis, respectively. We refer the reader to App.~\ref{AppPSBinning} for more details. Comparisons with Gaussian realisations are not included in this section, as these have the same power spectrum as the target lightcone. 

Figure~\ref{fig:PS_3D} shows a comparison between the 3D isotropic power spectrum of the target lightcone and that of the synthesised lightcones. The power spectrum of the synthesised lightcones closely follows that of the target lightcone, falling within $\pm {\sigma_s}$ on most scales. While it is difficult to estimate $\sigma_t^\text{oct}$ for the power spectrum, since the number of modes per bin differs for different data sizes, the averaged 3D power spectrum of the synthesised lightcones reproduces those of $d_t$ within 25$\%$ at most between .1 and 1 $h$Mpc$^{-1}$. Figures.~\ref{fig:PS_2D_XY_Comp} and~\ref{fig:PS_2D_XZ_Comp} show similar results for the 2D power spectra in the \(xy-\) and \(xz-\) planes. For these, a visible bias begins to appear at the smallest scales, and the syntheses reproduce the power spectra of $d_t$ to within 15$\%$ on average between 0.2 and 1 $h$Mpc$^{-1}$.

%A comparison of the 2D power spectra is shown in Figures \ref{fig:PS_2D_XY_Comp} and \ref{fig:PS_2D_XZ_Comp}. The mean 2D power spectra of the synthesised lightcones, on the $xy$ plane and on the $xz$ plane, can reproduce the 2D power spectrum of the target lightcone to within 15$\%$ between .2 and 1 $h$Mpc$^{-1}$, with the power spectra of the target lightcones falling within $1\sigma$ of the power spectra across the 30 lightcone realisations.

Fig.~\ref{fig:Cylin_PS_Comp} shows the comparison of the cylindrically averaged power spectra, where the synthesised lightcones are represented by the mean spectrum of the 30 realisations. The right panel shows the ratio between the cylindrically averaged power spectrum of the target light cone and that of the mean of the synthesised lightcones. While we see that the cylindrical power spectra are as a whole well reproduced, a notable feature of Fig. \ref{ratio} is the large ratio values at high $k_\parallel$ and $k_\perp$. These values at $k$ scales correspond to the bias we see at high $k$ in the power spectrum comparison in Fig. \ref{fig:PS_combined}, close to Nyquist frequencies. Apart from the bias observed at high $k$ modes, the cylindrical power spectra of the synthesised lightcones remain within $\pm 1.5 \tilde{\sigma}$ of the variance across the 30 realisations at all other scales.

Regarding the bias that appeared at high $k$ in the power spectra of the synthesised lightcones, particularly along the $z$ domain, it should be noted that this is not particularly a limitation of the $C^{11}$ and $C^{11'}$ power spectrum constraints. In fact, prior to the application of the inverse quantile transform, the power spectrum is recovered very well at these scales, although one could detect a very small amount of bias. However, this bias increases significantly when the transform is subsequently applied. This could be explained by the fact that the non-Gaussian structures, i.e. the statistical dependence between the different scales, are not properly reproduced at small scales. In this case, the highly non-linear inverse histogram transform could introduce such a bias. To address this issue, the $C^{11}$ and $C^{11'}$ power spectrum constraints could be added after this transform, but at the cost of additional computational time.

%\IH{In the power spectra comparisons shown, there appears to be some bias at high $k$ in the spectra of the synthesised lightcones, notably along the $z$ domain. Prior to the application of the inverse quantile transform, the recovery of the target lightcone at high-$k$ is in line with the recovery at lower $k$ with some divergence as the smallest scales, which are more difficult to recover with the generative model. Comparing the results before and after the inverse quantile transform suggests that the inverse exacerbates the bias at high $k$. 
%The generative models have an additional constraint on the power spectrum via \({C}^{11'}\). Since inverting the quantile transform exacerbates biases, one could also impose the constraint on \({C}^{11'}\) after the inverse transform to further constrain the power spectrum and reduce the bias post-inverse transform.}

\begin{comment}
\IH{I'm wondering if the bias is still from the fact that we haven't fully mitigated the periodicity? Given that xz is more affected by the bias, i think it must be there? If you remember I said that there is an anisotropic power that is introduced by the evolution, it could be that the anisotropic power component of the ps is not the same in both (the different redshift bands did look variance like) }
\end{comment}

\begin{figure}
    \centering
    \includegraphics[width=\linewidth]{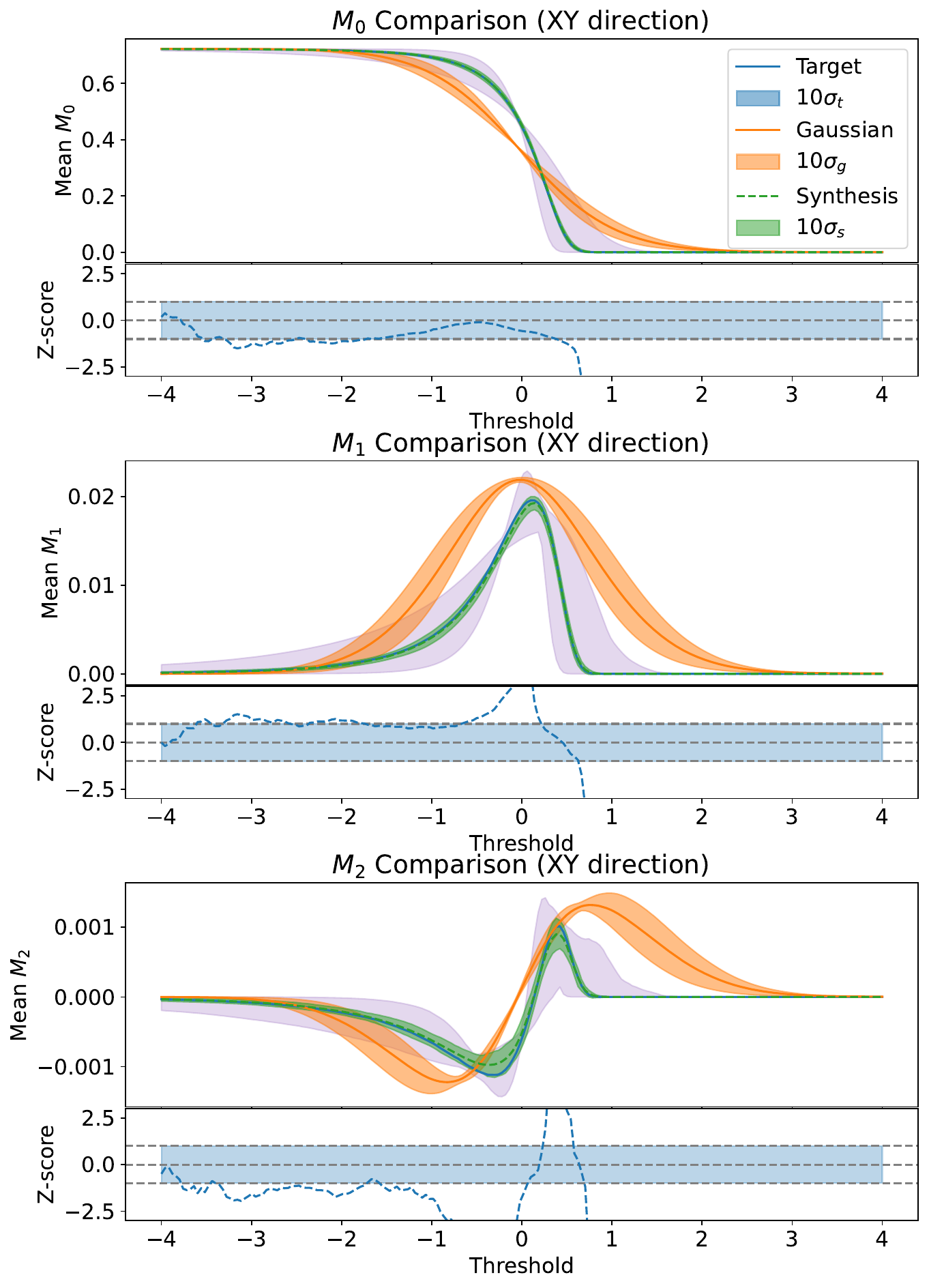}
\caption{Comparison of the Minkowski Functionals, applied to each frequency channel in the lightcone, between the synthesised lightcones (post-inverse quantile transform) and target lightcone. The mean Minkowski Functionals across the lightcone are shown, and the shaded 'error' regions are the standard deviation of the different Minkowski Functionals across the lightcone. From the bottom panel, $M_0$ and $M_1$ of the synthesised lightcones well recover the Minkowski Functionals of the target lightcone, lying within the octant variance, $\sigma_t^\text{oct}$, showing that it can reproduce the morphology of the target lightcone. The $M_2$ of the synthesised lightcone follows that of the target lightcone's $M_2$, and is within $\pm 2.5\sigma_t^\text{oct}$.}
\label{fig:Mink_Comp_XY}
\end{figure}

\begin{figure}
    \centering
    \includegraphics[width=\linewidth]{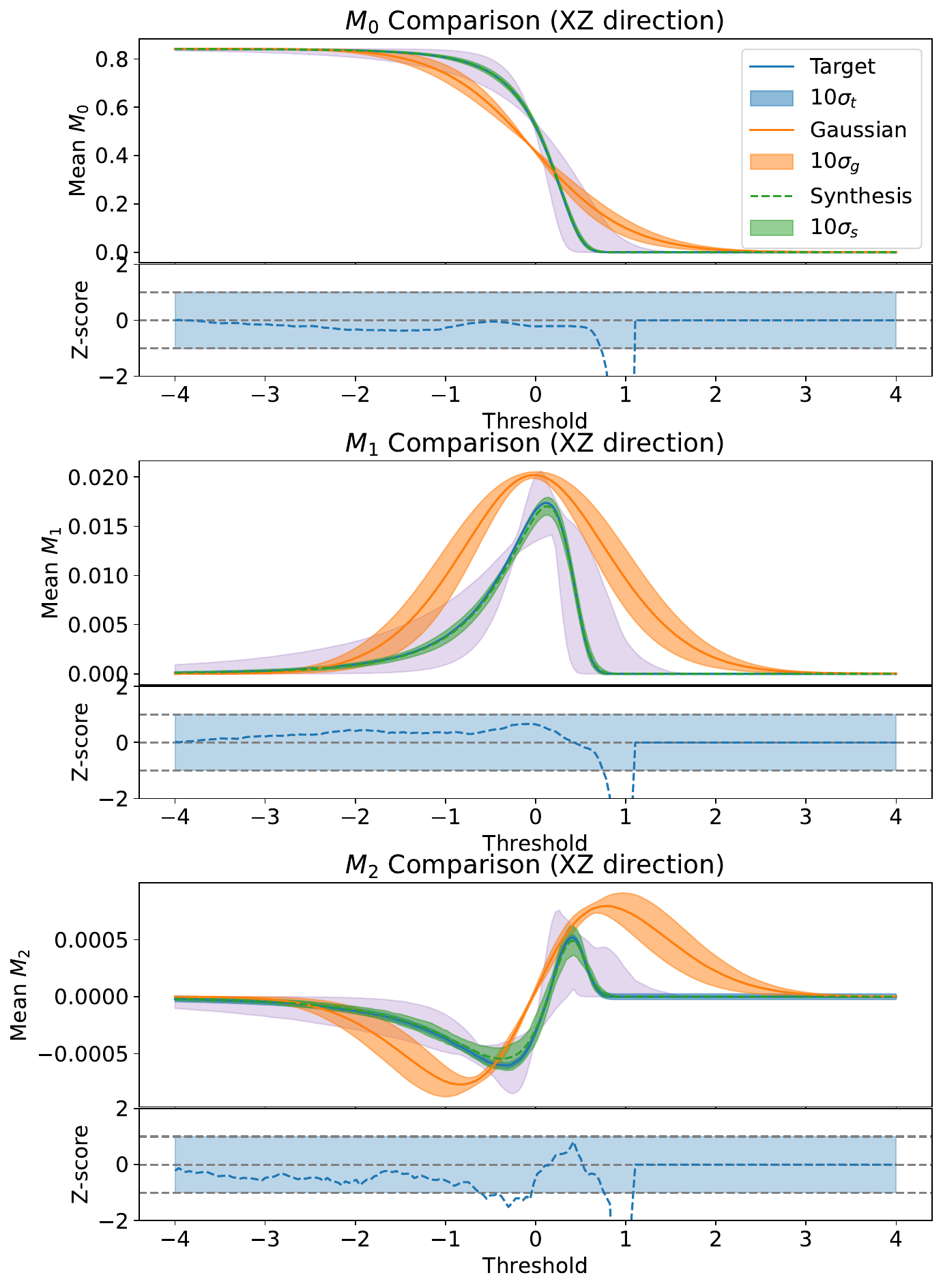}
\caption{Same as Fig. \ref{fig:Mink_Comp_XY}, except for the case of the $xz$-plane. The synthesised lightcones are able to reproduce the Functional of the $xz$-plane, and thus able to recover the morphology of the the evolving lightcone. The Minkowski Functionals of the synthesised lightcone lie within the octant variance, $\sigma_t^\text{oct}$ of the target lightcone.}
\label{fig:Mink_Comp_XZ}
\end{figure}

%%%%%%%%%%%%%%%%%%%%%%%%%%%%%%%%%%%%%%%%%%%%%%%%%%%%%%%%%%%%%%%%%%%%%%%%%%%%%%
\subsubsection{Minkowski Functionals}

Finally, we compare the Minkowski Functionals (MF) of the target and the synthesised lightcones, which were computed using the Minkowski Functionals function of the {\fontfamily{cmtt}\selectfont QuantImPy} Python package~\citep{2021:BoelensTchelepi}. As with the 2D isotropic power spectrum, in this paper we chose to compute the two-dimensional MFs on the \(xy\) or \(xz\) planes, which are then averaged over the \(z\) or \(y\) axes, respectively. There are three Minkowski Functionals (MFs), $M_0$, $M_1$, and $M_2$ \cite{1994:Mecke, 1997:Schmalzing}. These 2D MFs are computed by progressively increasing a threshold value, and evaluating the properties of the regions of values over these threshold: their total area ($M_0$) and perimeter ($M_1$), as well as their genus characteristic ($M_2$), that provides insight into the connectivity and complexity of ionised structures. 

To address the non-periodic boundary conditions of $d_t$, the MFs are in practice computed on the \(xy\) or \(xz\) planes after applying a complete symmetric padding to restore periodicity without significantly altering the underlying data (this results in some sharp edges at the initial boundary position). We also apply this symmetric padding when estimating the MFs of the syntheses, even if they have periodic boundary conditions, to enable the best possible comparison.

The MFs in the $xy$ and $xz$ planes are shown in Figures.~\ref{fig:Mink_Comp_XY} and~\ref{fig:Mink_Comp_XZ}, respectively. As the lightcone-to-lightcone standard deviations ($\sigma_t$, $\sigma_s$, and $\sigma_g$) of these statistics are very small, they have been multiplied by a factor of 10 in the plots.
These plots show that all three MFs of the synthesised lightcones closely resemble those of $d_t$ in both $xy$ and $xz$ planes. This indicates that the ST generative models effectively reproduce the morphological features of the data. However, it is worth noting that while the difference between these mean statistics mostly stays within $\pm \sigma_t^\text{oct}$ for the $xz$-MFs, this is not entirely the case for the $xy$-MFs, for which clear deviations are particularly visible for $M_2$. We interpret this as the first limitation of the ST generative model, arising from its difficulty in reproducing the sharp, high-contrast and highly textured circular features present in the target lightcone, as can be seen in Fig.~\ref{fig:LoReLi_Comparison_XY_Lightcone}. Overall, for both $xy$ and $xz$ cases, the $M_0$, $M_1$, and $M_2$ statistics are still reproduced within $1\%$, $10\%$, and $15\%$, respectively.

Finally, it is interesting to note that the synthesised data have a lightcone-to-lightcone standard deviation, $\sigma_s$, which is a few times smaller than the standard deviation $\sigma_t$ estimated from $d_t$. This is in addition to the fact that Gaussian models have, on the contrary, a considerably larger variance. This could indicate a second limitation of ST-generative models. However, care should be taken when interpreting this result, since estimating $\sigma_t$ from $\sigma_t^\text{oct}$ could lead to a substantial underestimation of this variance, which could be larger between lightcone than within them.

\section{Conclusion}
\label{Conclusion}

\begin{comment}
The 21cm signal from the Epoch of Reionization (EoR) is observed by radio interferometers as a lightcone - which encompasses a (2D) sky plane and a (1D) redshift domain that captures the evolution of the 21cm signal. These observations are contaminated by foregrounds that are several orders of magnitude stronger than the EoR signal radio interferometers aim to detect. Current foreground removal techniques are optimised to recover the power spectrum, but they fail to recover the non-Gaussian properties of the signal. Recent works focused on the interstellar medium have shown that scattering transform-based component separation methods can effectively extract these non-Gaussian fields from foreground-contaminated data. Since existing component methods rely on 2D scattering transforms, to adapt them from EoR studies we extend these transforms to 3D to summarise lightcones.
\end{comment}

The 21cm signal from the Epoch of Reionisation (EoR) is observed by radio interferometers as a lightcone that includes a (2D) sky plane and a (1D) redshift domain that captures the evolution of the 21cm signal. These observations are contaminated by foregrounds that are several orders of magnitude stronger than the EoR signal that radio interferometers are designed to detect. Recently developed component separation methods rely on maximum entropy generative models built from scattering transforms. In this paper we explore the feasibility of generative models of EoR lightcones built from 3D scattering transforms.

These 3D scattering transforms are built from a wavelet set constructed by taking the tensor product between a 1D wavelet set defined in $k_z$ and a 2D wavelet set defined in the $k_x - k_y$ plane. We demonstrated the performance of these generative models by applying them to a single simulated EoR lightcone from the LoReLi II database. The generative model was validated by comparing several realisations of the synthesised lightcone to the original target lightcone using a number of independent statistical estimators, including the histogram, 3D and 2D power spectra, cylindrically averaged power spectrum, and Minkowski Functionals.

The results showed that the synthesised lightcones reproduce very well the key non-Gaussian properties and morphology of the target lightcone, including its characteristic skewed histogram, redshift-dependent structures, and morphological complexity as probed by the Minkowski Functionals. These statistics were well reproduced within the expected sample variance estimated from the data.

Beyond the overall validation, some specific current limitations have arisen:
\begin{itemize}
    %\item An important practical consideration in this work is the treatment of the non-periodic boundary conditions inherent in simulated EoR lightcones, particularly along the redshift axis due to cosmic evolution. To mitigate the boundary effects introduced by the Fourier-based convolutions used in the scattering transform, we have restricted the computation of the scattering statistics to a central sub-volume of the lightcone. While this approach ensures that the estimated statistics are not biased by boundary artefacts, it effectively reduces the usable volume by approximately one order of magnitude. This reduced data volume introduces additional sampling variance and limits the amount of statistical information available to constrain the generative model, which may contribute to some of the observed biases in the model outputs.
    
    \item Although the synthesised lightcones were able to reproduce the power spectra of the target lightcones over a wide range of scales, there were clear biases that occurred at high $k$ scales (small scales), particularly along the redshift axis, suggesting limitations of the generative model when combined with the inverse quantile transformation. However, this could be mitigated by additional post-inverse corrections or by refining the treatment of scales not constrained by the scattering transforms.

    \item The synthesised lightcone Minkowski Functionals did not fully reproduce the $M_2$ of the target lightcone. This suggests that the generative model struggles to produce sharp, high-contrast circular features in the target lightcone, such as ionised regions. This issue could probably be solved by using more sophisticated scattering transform representation, such as the complete Scattering Spectra statistics~\citep{2023:ChengMorelAllys}, but at an additional computational cost.

    \item The synthesised lightcones also showed a large variety of realisations, although the variance between realisations appeared to be slightly overestimated compared to the variance estimated over the octants of the target lightcone.  However, to investigate this point in detail, more simulated data would be required to properly estimate the lightcone-to-lightcone variability.
    %This may reflect either an intrinsic limitation of the model or an artefact of the approximate variance estimation method.
\end{itemize}

Overall, this work provides the first proof-of-concept for the application of 3D scattering transform-based generative models to EoR lightcones. The results of this work pave the way for the development of a statistical component separation based on the 3D scattering transforms developed statistics to separate the EoR signal from a realistic forward modelled lightcone with foregrounds and noise.

%%%%%%%%%%%%%%%%%%%%%%%%%%%%%%%%%%%%%%%%%%%%%%%%%%%%%%%%%%%%%%%%%%%%%%%%%%%%%%
\begin{acknowledgements}
The post-doctoral contract of Ian Hothi was funded by Sorbonne Université in the framework of the Initiative Physique des Infinis (IDEX SUPER) and through the overheads of Edith Falgarone's MIST ERC. We would like to thank Florent Mertens for their useful discussions and input. The Loreli project was provided with computer and storage resources by GENCI at TGCC thanks to the grants 2022-A0130413759 and 2023- A0150413759 on the supercomputer Joliot Curie’s ROME partition. This work was granted access to the HPC resources of MesoPSL financed by the Region Ile de France and the project Equip@Meso (reference ANR-10-EQPX-29-01) of the programme Investissements d’Avenir supervised by the Agence Nationale pour la Recherche.

This research made use of astropy, a community-developed core Python package for astronomy \citep{Astropy}; scipy, a Python-based ecosystem of open-source software for mathematics, science, and engineering \citep{Scipy} - including numpy \citep{numpy}; matplotlib, a Python library for publication quality graphics \citep{matplotlib}.
\end{acknowledgements}

%%%%%%%%%%%%%%%%%%%%%%%%%%%%%%%%%%%%%%%%%%%%%%%%%%%%%%%%%%%%%%%%%%%%%%%%%%%%%%
%%%%%%%%%%%%%%%%%%%%%%%%%%%%%%%%%%%%%%%%%%%%%%%%%%%%%%%%%%%%%%%%%%%%%%%%%%%%%%
\appendix
\section{Power Spectrum Binning}
\label{AppPSBinning}

The dimensionless power spectrum of the 21cm brightness temperature \(I\) is defined as
\begin{equation}
\Delta_i^2(k) = \frac{A_k}{2\pi_i^2} \int \bigl|\tilde{I}(\vec{k}) \cdot W_i(\vec{k})\bigr|^2 \, d\vec{k},
\end{equation}
where $\vec{k} ~(= k_x,k_y,k_z)$ is the wavenumber and  \(W_i(\vec{k})\) is the spectral window function that defines the \(i\)th bin. We choose the spectral window function to be a top-hat function defined as, 
\begin{equation}
    W_i(\vec{k}) = \begin{cases}
1, & \text{if } \vec{k}_i \leq \vec{k} < \vec{k}_{i+1}\\[1mm]
0, & \text{otherwise}
\end{cases}
\end{equation}
where $\vec{k}_i$ defines the lower wavenumber limit of the $i^{th}$ bin. These window functions are assumed to be isotropic, so \(W(\vec{k}) = W(k)\) with \(k = \lvert \vec{k}\rvert\). For the 3D power spectrum \(A_k = k^3\) and for the 2D power spectrum \(A_k = k^2\).

Before estimating the power spectrum, it is essential to account for the non-periodic boundary conditions of the target lightcone. We address this by applying a Blackman-Harris window function to the data prior to the Fourier transform. Specifically, for the 3D power spectrum, we apply the 1D Blackman-Harris window function separately along each axis of the 3D lightcone. Similarly, for the 2D power spectrum, this window is applied independently along each axis of the respective 2D field.
%The 1D Blackman-Harris window function applied along each axis is given by
%\begin{equation}
%    w(x) = a_0 - a_1 \cos\left(\frac{2\pi x}{L}\right) + a_2 \cos\left(\frac{4\pi x}{L}\right) - a_3 \cos\left(\frac{6\pi x}{L}\right),
%\end{equation}
%where \(x\) is the coordinate along the axis of length \(L\) and the coefficients are \(a_0 = 0.35875\), \(a_1 = 0.48829\), \(a_2 = 0.14128\) and \(a_3 = 0.01168\).
This windowing procedure is applied consistently to both the target \(d_t\) and synthesised \(\tilde{d}_i\) lightcones; although the \(\tilde{d}_i\) have periodic boundary conditions, this application ensures consistent treatment of boundary effects.

In addition, we also restrict the power spectra estimation to the range of scales that are well constrained by the generative model, so that the range of $k$ modes that are binned is limited to $\kappa_{low} \leq k \leq \kappa_{high}$ (see Section \ref{limiting_scales} for definitions).
\color{black}
The dimensionless cylindrically averaged power spectrum, which bins the \(k_\parallel\) and \(k_\perp\) domains separately, is defined as
\begin{equation}
\Delta_i^2(k_\parallel, k_\perp) = \frac{A_k}{2\pi_i^2} \int \bigl|\tilde{I}(\vec{k_\parallel}, \vec{k_\perp}) \cdot W_i(\vec{k_\parallel}, \vec{k_\perp})\bigr|^2 \, d\vec{k_\parallel}\,d\vec{k_\perp},
\end{equation}
where \(A_k = k^3\), for $k = \sqrt{\vec{k_\parallel}^2 +  \vec{k_\perp}^2}$. We again assume an isotropic top-hat function, $W_i({k_\parallel}, {k_\perp})$, for \(k_\parallel = \lvert \vec{k_\parallel}\rvert\) and \(k_\perp = \lvert \vec{k_\perp}\rvert\), yielding
\begin{equation}
    W_i(k_\parallel, k_\perp) =
\begin{cases}
1, & \text{if } k_{\parallel,i} \le k_\parallel < k_{\parallel,i+1} \;\&\;k_{\perp,i} \le k_\perp < k_{\perp,i+1}\\[1mm]
0, & \text{otherwise}
\end{cases}
\end{equation}
where $k_{\parallel,i}$ defines the lower wavenumber limit of the $i^{th}$ along the $k_{\parallel}$ domain, and $k_{\perp,i}$ defines the lower wavenumber limit of the $i^{th}$ along the $k_{\perp}$ domain.
%%%%%%%%%%%%%%%%%%%%%%%%%%%%%%%%%%%%%%%%%%%%%%%%%%%%%%%%%%%%%%%%%%%%%%%%%%%%%%
\bibliography{cit} 

\end{document}